\documentclass[usenatbib, usegraphicx]{mn2e}
\newcommand{\Lyalpha}{Ly$\alpha$}
\newcommand{\Lybeta}{Ly$\beta$}
\newcommand{\Lygamma}{Ly$\gamma$}
\newcommand{\Mdot}{\ensuremath{\dot{M}}}
\newcommand{\MdotA}{\ensuremath{\Mdot_\mathrm{A}}}
\newcommand{\MdotB}{\ensuremath{\Mdot_\mathrm{B}}}
\newcommand{\Msolpy}{\ensuremath{\mbox{~M}_\odot\mbox{~yr}^{-1}}}
\newcommand{\Rsol}{\ensuremath{\mbox{~R}_\odot}}
\newcommand{\nel}{\ensuremath{n_{\mathrm{e}}}}
\newcommand{\nH}{\ensuremath{n_{\mathrm{H}}}}
\newcommand{\R}{\ensuremath{{\mathcal R}}}
\newcommand{\vA}{\ensuremath{v_\mathrm{A}}}
\newcommand{\vB}{\ensuremath{v_\mathrm{B}}}
\newcommand{\angstrom}{\ensuremath{\mbox{~\AA}}}
\newcommand{\cm}{\ensuremath{\mbox{~cm}}}
\newcommand{\kelvin}{\ensuremath{\mbox{~K}}}
\newcommand{\kev}{\ensuremath{\mbox{~keV}}}
\newcommand{\kmps}{\ensuremath{\mbox{~km~s}^{-1}}}
\newcommand{\ks}{\ensuremath{\mbox{~ks}}}
\newcommand{\OVIII}{O\,\textsc{viii}}
\newcommand{\NeX}{Ne\,\textsc{x}}
\newcommand{\MgXII}{Mg\,\textsc{xii}}
\newcommand{\SiXIV}{Si\,\textsc{xiv}}
\newcommand{\SXVI}{S\,\textsc{xvi}}
\newcommand{\asca}{\textit{ASCA}}
\newcommand{\chandra}{\textit{Chandra}}
\newcommand{\einstein}{\textit{Einstein}}
\newcommand{\xmm}{\textit{XMM--Newton}}

\title{Theoretical X-ray Line Profiles from Colliding Wind Binaries}
\author[D. B. Henley, I. R. Stevens \& J. M. Pittard]
  {David B. Henley,$^1$\thanks{Email: dbh@star.sr.bham.ac.uk} Ian R. Stevens$^1$ and Julian M. Pittard$^2$  \\
   $^1$School of Physics and Astronomy, University of Birmingham, Edgbaston, Birmingham B15 2TT \\
   $^2$Department of Physics and Astronomy, University of Leeds, Woodhouse Lane, Leeds LS2 9JT}

\begin{document}

\maketitle

\begin{abstract}

We present theoretical X-ray line profiles from a range of model colliding wind systems. In particular, we investigate
the effects of varying the stellar mass-loss rates, the wind speeds, and the viewing orientation. We find
that a wide range of theoretical line
profile shapes is possible, varying with orbital inclination and phase. At or near conjunction, the lines have
approximately Gaussian profiles, with small widths ($\mathrm{HWHM} \sim 0.1 v_\infty$) and definite blue- or 
redshifts (depending on whether the star with the weaker wind is in front or behind). When the system is viewed at 
quadrature, the lines are generally much broader ($\mathrm{HWHM} \sim v_\infty$), flat-topped and unshifted.
Local absorption can have a major effect on the observed profiles -- in systems with mass-loss rates 
of a few times $10^{-6} \Msolpy$ the lower energy lines ($E \la 1 \kev$) are particularly affected. This generally 
results in blueward-skewed profiles, especially when the system is viewed through the dense wind of the primary. 
The orbital variation of the line widths and shifts is reduced in a low inclination
binary. The extreme case is a binary with $i = 0\degr$, for which we would expect no line profile variation.

\end{abstract}


%%%%%%%%%%%%%%%%%%%%%%%%%%%%%%%%%%%%%%%%%%%%%%%%%%%%%%%%%%%%%%%%%%%%%%%%%%%%%%%%%%%%%%%%%%%%%%%%%%%%%%%%%%%%%%%%%%%%%%%%

\section{Introduction}
\label{sec:Introduction}

\citet{cherepashchuk76} and \citet{prilutskii76} proposed that the collision of the winds in
a binary of early-type stars should produce copious quantities of X-ray-producing gas.
\citet{pollock87} found that among the 48 Wolf--Rayet stars observed with \einstein,
the binary systems tended to be a few times brighter in X-rays than single stars. There is a similar
trend among the O stars observed with \einstein\ \citep{chlebowski89b}. This provided
early observational evidence for colliding wind X-ray emission.

Colliding wind binaries (CWBs) have also been studied extensively with numerical
hydrodynamical simulations \citep*[e.g.][]{luo90,stevens92,pittard97,pittard98b}.
These works investigated the broad-band X-ray properties, since the satellites
in operation at that time had only poor spectral resolution 
(e.g. $E / \Delta E \sim 20$ for the \asca\ SIS). Nevertheless, by comparing the variable
\asca\ spectrum of $\gamma^2$ Velorum with hydrodynamical models of the system,
\citet{stevens96} were able to confirm that the system is indeed a CWB, and were also
able to place constraints on the stars' mass-loss rates and wind velocities. This technique
has more recently been applied to the high-resolution \chandra\ grating spectrum of $\eta$ Carinae
\citep{pittard02a}.

The unprecedented spectral resolution offered by the \chandra\ LETGS and HETGS and the \xmm\ RGS 
($E / \Delta E \sim 100$--1000) gives us the opportunity to study in detail the profiles of individual
X-ray lines from CWBs, enabling us to probe the very hot shocked gas in the wind-wind interaction regions.
Studies of X-ray line profiles from single early-type stars have already been used to test the standard wind-shock
model of X-ray production, in which instabilities due to the line-driving force give rise to shocks distributed 
throughout the stellar wind \citep*{lucy80,lucy82,owocki88,feldmeier97a}. Line profile calculations based 
upon such models (with X-ray production effectively distributed uniformly throughout a steady-state
or accelerating stellar wind) generally predict broad, blueshifted and blue-skewed line profiles 
\citep{ignace01a,owocki01}. However, blueshifted lines have only been observed from $\zeta$ Puppis
\citep{kahn01,cassinelli01a}, while the lines observed from $\delta$ Orionis \citep{miller02b} and 
$\tau$ Scorpii \citep{cohen03} are far narrower than is expected from the standard wind shock model.

An analogous study of X-ray line profiles from CWBs has not yet been made. Only relatively few observations 
of CWBs (or suspected CWBs) with high resolution spectrometers have been published; the best examples are 
$\gamma^2$ Vel \citep{skinner01} and $\eta$ Car (\citealt{corcoran01b,pittard02a}; \citealt*{leutenegger03}).
In these works, very little attention has been paid to the line profile shapes. The \chandra\ HETGS spectrum of
$\gamma^2$ Vel is dominated by emission from H- and He-like ions of Ne, Mg, Si and S. The lines are broader than
the expected thermal widths ($\mathrm{HWHM} \approx 500$--700\kmps), and there are no significant wavelength 
shifts \citep{skinner01}. The lines in the \chandra\ HETGS spectrum of $\eta$ Car (from H-like ions of Mg, Si 
and S and He-like ions of Si, S, Ca and Fe) are also not significantly shifted \citep{corcoran01b}. No
mention is made of the widths of the lines. Similarly, no mention is made of line widths or shifts in the 
\xmm\ RGS spectrum of $\eta$ Car \citep{leutenegger03}. However, there is evidence for orbital variation of
line shifts in the \chandra\ HETGS spectrum of WR 140 (Pollock et al., in prep.) -- this will be discussed
further in \S\ref{sec:Discussion}.

We present here results of theoretical calculations of X-ray line profiles from such systems
The model is described in \S\ref{sec:Model}. The calculated profiles are presented in \S\ref{sec:Results}, 
in particular showing how the profiles vary with wind parameters and the viewing orientation. The results are 
discussed in \S\ref{sec:Discussion} and summarized in \S\ref{sec:Summary}. The line profiles presented here
will ultimately be compared with X-ray spectra of CWBs, and potentially offer another
tool for placing constraints on the wind parameters of such systems, as well as 
providing new insights into the structure of the X-ray emitting regions.

%%%%%%%%%%%%%%%%%%%%%%%%%%%%%%%%%%%%%%%%%%%%%%%%%%%%%%%%%%%%%%%%%%%%%%%%%%%%%%%%%%%%%%%%%%%%%%%%%%%%%%%%%%%%%%%%%%%%%%%%

\section{The model}
\label{sec:Model}

\subsection{Hydrodynamic simulations}
\label{subsec:Hydrodynamics}

\begin{figure*}
  \includegraphics[width=8cm]{md56_fig01a.eps}
  \hspace{1cm}
  \includegraphics[width=8cm]{md56_fig01b.eps}
  \caption{Density (left) and temperature (right) maps from a hydrodynamic simulation of a colliding wind binary
  with $\R = \sqrt{5}$ ($\MdotA = 5 \times 10^{-6} \Msolpy$, $\MdotB = 1 \times 10^{-6} \Msolpy$,
  $\vA = \vB = 2000 \kmps$).}
  \label{fig:DensityandTemperatureMaps}
\end{figure*}

The hydrodynamic simulations of CWBs were carried out using \textsc{vh-1}, an
implementation of the piecewise parabolic method \citep[PPM;][]{colella84} developed by John
Blondin and co-workers at the University of Virginia\footnote{http://wonka.physics.ncsu.edu/pub/VH-1}.

The CWBs are characterized by the wind momentum ratio \R, given by

\begin{equation}
\R = \sqrt{\frac{\MdotA \vA}{\MdotB \vB}}
\label{eq:WindMomentumRatuio}
\end{equation}

\noindent
where \MdotA\ and \MdotB\ are the mass-loss rates of the stars and \vA\ and \vB\ are the terminal velocities
of the winds \citep{stevens92}.
By definition, star A (the primary) is the one with the more powerful wind (i.e. $\R \ge 1$).

Modelling the flow in a CWB is a three-dimensional problem. However, except in
very close binaries, the orbital velocity of the stars is much less than the terminal velocity of
the winds. Since we are not modelling close binaries, Coriolis forces are assumed to be negligible,
and the problem reduces to a two-dimensional axisymmetric flow. This greatly reduces the amount of
computer time needed to model the system.

It is important to note that the goal of this work is to gain insights into where different X-ray lines
originate in the wind-wind collision zone, and how the profiles are affected by different wind parameters.
As a result, rather than modelling specific CWBs, we have chosen wind parameters that
are representative of typical systems. In particular, the stellar separation is fixed at
$2 \times 10^{13} \cm$ (287\Rsol) , mass-loss rates are typically $\sim 10^{-6}$--$10^{-5} \Msolpy$, and wind
speeds are typically 2000\kmps. 

For these wind parameters the winds can be assumed to be adiabatic, because the cooling time
$t_\mathrm{cool}$ is large compared with $t_\mathrm{esc}$, the escape time from the shock region;
i.e. the cooling parameter $\chi = t_\mathrm{cool} / t_\mathrm{esc}$
\citep{stevens92} is large. A few simulations were carried out with radiative cooling included 
self-consistently. The inclusion of cooling did not significantly affect the
results, and so in general it was not included at all.

A further assumption is that the winds are at their terminal velocities ($v_\infty$) when they collide. This
assumption is valid for wide systems, since the winds of early-type stars typically reach their
terminal velocities within a few stellar radii of the stellar surface. This assumption may be relaxed
in the future for models of specific close binary systems, using the method described in
\citet{pittard98b}.

Fig.~\ref{fig:DensityandTemperatureMaps} is an example snapshot from one of the hydrodynamical simulations,
showing the temperature and density structure of a CWB with $\R = \sqrt{5}$
($\MdotA = 5 \times 10^{-6} \Msolpy$, $\MdotB = 1 \times 10^{-6} \Msolpy$, $\vA = \vB = 2000 \kmps$).
For these wind parameters, the cooling parameters \citep{stevens92} are $\chi_\mathrm{A} = 4.2$ and
$\chi_\mathrm{B} = 11.2$, and so the winds can indeed be assumed to be adiabatic. Note also that instabilities
can be seen in the shocked gas, which may be due to either a difference in flow velocities along the two sides of the
contact discontinuity between the two winds \citep{stevens92} or numerical effects near the line of centres.

\subsection{Line profile calculations}
\label{subsec:LineProfileCalculations}
The method used for calculating X-ray line profiles from the hydrodynamic simulations simply involves
calculating the line profile for each point on the grid and summing over the whole grid.

Thermal Doppler broadening is the dominant effect, and so each grid point is assumed
to produce a Gaussian line 
profile whose height and width depend on the temperature of the gas at that point,
and whose centre is blue- or redshifted according to the line-of-sight velocity of the gas, i.e.

\begin{equation}
I^i(v) = I_0^i \exp \left( \frac{-m (v-v_\mathrm{los}^i)^2}{2kT^i} \right)
\label{eq:GaussianProfile}
\end{equation}

\noindent
where $I^i(v)\mathrm{d}v$ is the intensity of the line from the $i$th grid point
between a velocity shift of $v$ and $v + \mathrm{d}v$,
$m$ is the ion mass, $T^i$ is the gas temperature, and $v_\mathrm{los}^i$ is the line-of-sight velocity
of the gas.

The height of the Gaussian $I_0^i$ is calculated using the list of spectral lines available from the
\textsc{spex} website\footnote{http://www.sron.nl/divisions/hea/spex/version1.10/line/ \\ index.html}.
This list gives the emissivity $Q$ of a few thousand UV and X-ray lines over a range of temperatures.
For an arbitrary temperature, the emissivity is calculated by interpolating between the tabulated
values. If the temperature lies outside the range of temperatures for which there are data, the emissivity
is assumed to be zero and the grid point is skipped over. The total line luminosity $L_\mathrm{line}^i$
for the $i$th grid point is obtained by multiplying these emissivities by $\nel^i \nH^i V^i$, where 
$\nel^i$ and $\nH^i$ are the electron and hydrogen number densities and $V^i$ is the volume of the emitting gas 
at the grid point. $L_\mathrm{line}^i$ is equal to the integral of equation (\ref{eq:GaussianProfile}) 
over all $v$, and so

\begin{equation}
I_0^i = Q(T^i) \nel^i \nH^i V^i \sqrt{\frac{m}{2 \pi kT^i}}
\label{eq:GaussianHeight}
\end{equation}

Two major assumptions have to be made in order to use the data in the \textsc{spex} list of lines. Firstly, the
plasma is assumed to be in collisional ionization equilibrium, and secondly both winds are assumed to have
solar abundances. The latter is a valid assumption for O+O binaries, but not for WR+O binaries, in which
the Wolf--Rayet wind has significantly non-solar abundances. This assumption will be relaxed in the future
when specific WR+O binaries are modelled.

The line profile produced by each grid point is then attenuated by photoabsorption.
The absorbing column to each emitting grid point is 
calculated by integrating the density numerically along the line-of-sight. The density is either 
obtained directly from the hydrodynamic grid, or by extrapolation when off the grid. The opacities 
used were calculated for a solar abundance plasma using 
\textsc{xstar}\footnote{http://heasarc.gsfc.nasa.gov/docs/software/lheasoft/xstar/ \\ xstar.html}.
For a given line, the opacity is assumed to be constant with wavelength across the width of the line profile.
Most of the absorption takes place in the cool, unshocked winds surrounding the wind-wind collision region,
which are both assumed to have a temperature of $3 \times 10^4 \kelvin$. However, opacities have also been calculated
for higher temperatures ($10^5$, $10^6$, $10^7$ and $10^8 \kelvin$). The opacity of hot, shocked gas of arbitrary
temperature in the wind-wind collision region is taken into account by interpolating between these values,
though in fact this gas makes only a small contribution to the overall absorption.
When calculating the contribution of gas off the grid to the absorption, we assume its temperature is equal to the 
gas temperature at the point on the line-of-sight where we leave the grid. This tends to underestimate the absorption, 
as it sometimes assumes that gas off the grid is hotter than it actually is. The alternative is to assume that all the gas 
off the grid is cold ($T = 3 \times 10^4 \kelvin$), which would tend to overestimate the absorption. However, we 
have found that there is generally very little difference in the resulting profiles calculated by the two methods.

The resulting line profile from the CWB is the sum of all the profiles calculated
by the above method for the entire grid. Results for a range of CWB parameters are shown in the following section.

%%%%%%%%%%%%%%%%%%%%%%%%%%%%%%%%%%%%%%%%%%%%%%%%%%%%%%%%%%%%%%%%%%%%%%%%%%%%%%%%%%%%%%%%%%%%%%%%%%%%%%%%%%%%%%%%%%%%%%%%

\begin{table}
\caption{The wavelengths of the \Lyalpha\ lines used in this work (taken from \textsc{spex} list of lines).
$\lambda_1$ and $\lambda_2$ are the wavelengths of the brighter and weaker components, respectively.
$\Delta v$ is the difference in wavelengths expressed as a velocity shift
($\Delta v = c (\lambda_2 - \lambda_1) / \lambda_1$).}
\begin{tabular}{lr@{.}lr@{.}lc}
\hline
Ion     & \multicolumn{2}{c}{$\lambda_1$ (\AA)} & \multicolumn{2}{c}{$\lambda_2$ (\AA)} & $\Delta v$ (\kmps) \\
\hline
\OVIII  & 18&971 & 18&977 &  95 \\
\NeX    & 12&132 & 12&137 & 124 \\
\MgXII  &  8&419 &  8&424 & 178 \\
\SiXIV  &  6&181 &  6&187 & 291 \\
\SXVI   &  4&729 &  4&733 & 254 \\
\hline
\end{tabular}
\label{table:LineWavelengths}
\end{table}

%%%%%%%%%%%%%%%%%%%%%%%%%%%%%%%%%%%%%%%%%%%%%%%%%%%%%%%%%%%%%%%%%%%%%%%%%%%%%%%%%%%%%%%%%%%%%%%%%%%%%%%%%%%%%%%%%%%%%%%%

\section{Results}
\label{sec:Results}
Line profiles were calculated for the \Lyalpha\ lines from the abundant elements
O, Ne, Mg, Si and S. These lines are in fact closely spaced doublets; both components
were included in the calculations, but other nearby lines were ignored (as their emissivities
are much lower than those of the \Lyalpha\ lines). The wavelengths of the lines are shown in
Table~\ref{table:LineWavelengths}.

The O to S \Lyalpha\ lines were chosen because they feature strongly in the X-ray grating spectra of
many early-type stars \citep{schulz00a,waldron01,kahn01,cassinelli01a,miller02b}.
They are also insensitive to the plasma density, unlike the \textit{fir} 
(forbidden-intercombination-resonance) triplets from He-like ions. This greatly simplifies the calculations.

%%%%%%%%%%%%%%%%%%%%%%%%%%%%%%%%%%%%%%%%%%%%%%%%%%%%%%%%%%%%%%%%%%%%%%%%%%%%%%%%%%%%%%%%%%%%%%%%%%%%%%%%%%%%%%%%%%%%%%%%

\subsection{Location of the line-emitting plasma}
\label{subsec:PlasmaLocation}

\begin{figure}
  \centering
  \includegraphics[width=8cm]{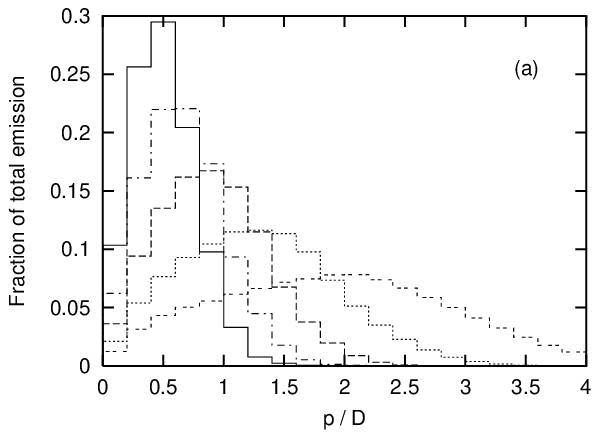}
  \includegraphics[width=8cm]{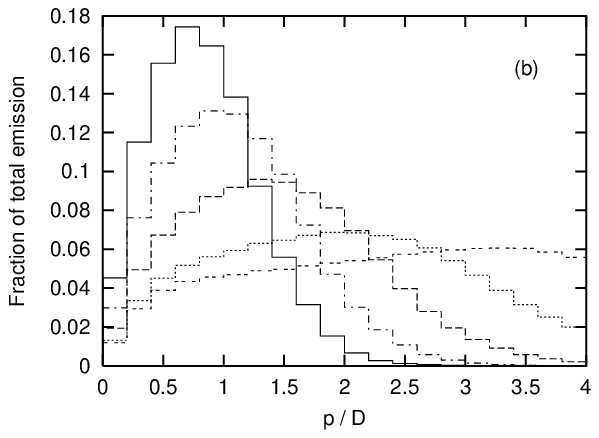}
  \caption{The fraction of the total line emission as a function of distance $p$ from the line of centres,
  for the \Lyalpha\ lines from \OVIII\ (double-dashed), \NeX\ (dotted), \MgXII\ (dashed),
  \SiXIV\ (dot-dashed) and \SXVI\ (solid). The mass-loss rates are $\MdotA = 5 \times 10^{-6} \Msolpy$ and
  $\MdotB = 1 \times 10^{-6} \Msolpy$. The wind speeds are (a) $\vA = \vB = 2000 \kmps$,
  (b) $\vA = \vB = 3000 \kmps$. The stellar separation $D = 2 \times 10^{13} \cm$.}
  \label{fig:ZoneEmission}
\end{figure}

Fig.~\ref{fig:ZoneEmission} shows the fraction of the total unabsorbed line emission
as a function of cylindrical distance $p$ from the line of centres for each of five lines
(i.e. the figure shows the fractional line luminosity from a series of thin cylindrical shells
whose symmetry axes lie along the line of centres). The results in 
Fig.~\ref{fig:ZoneEmission}(a) are from a simulation with 
$\MdotA = 5 \times 10^{-6} \Msolpy$, $\MdotB = 1 \times 10^{-6} \Msolpy$
and $\vA = \vB = 2000 \kmps$. The results in Fig.~\ref{fig:ZoneEmission}(b) are from a
simulation with the same mass-loss rates, but with higher wind velocities (3000\kmps).
The simulations were carried out on an $8 \times 10^{13} \cm$ square grid with a stellar separation 
$D = 2 \times 10^{13} \cm$. In all cases, the values of $p$ are normalized to $D$.

In both cases, one can clearly see that most of the \Lyalpha\ emission from the lighter elements (e.g. O, Ne) 
originates a few times $10^{13} \cm$ from the line of centres, whereas that from heavier elements (e.g. S) 
originates much further in ($\sim 1 \times 10^{13} \cm$ from the line of centres).
This is exactly as expected, because the temperature of the plasma in the wind-wind collision region 
falls off with increasing $p$. In the inner regions it is too hot for \OVIII\ and \NeX\ to exist in large
amounts (these elements are mostly fully ionized), whereas in the outer regions it is too cool to have 
significant amounts of H-like ions of heavier elements.

The main difference between Figs.~\ref{fig:ZoneEmission}(a) and \ref{fig:ZoneEmission}(b) is that
for the system with higher wind velocities, the line emission tends to originate from larger
values of $p$. This is because the higher wind velocity results in a higher post-shock temperature.

For all the lines, the radius at which the emission peaks is smaller than the radius at which the
plasma temperature equals the temperature of maximum emissivity. This is because the emissivity $Q$ of the lines varies
fairly slowly with $p$. However, the $\nel^i \nH^i V^i$ term in equation (\ref{eq:GaussianHeight}) varies as approximately
$p^{-2}$, except near the line of centres, where it turns over and tends to zero. As a result the emission originates 
from further in than one might na\"{\i}vely expect from the temperature alone.

%%%%%%%%%%%%%%%%%%%%%%%%%%%%%%%%%%%%%%%%%%%%%%%%%%%%%%%%%%%%%%%%%%%%%%%%%%%%%%%%%%%%%%%%%%%%%%%%%%%%%%%%%%%%%%%%%%%%%%%%

\begin{figure*}
\begin{tabular}{cccc}
& $\R = 1$ & $\R = \sqrt{5}$ & $\R = 3$ \\
\raisebox{1.5cm}{$\theta = 0\degr$}
	& \includegraphics[width=5cm]{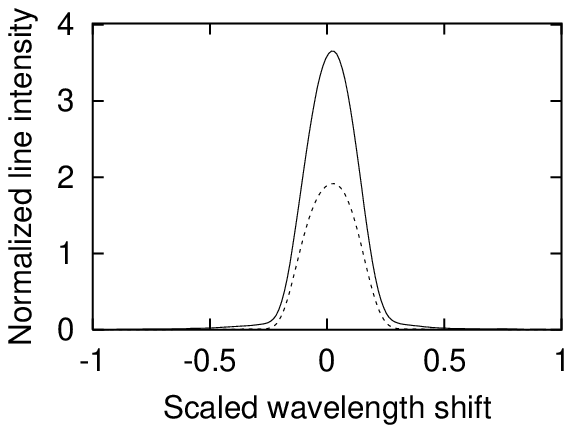} 
	& \includegraphics[width=5cm]{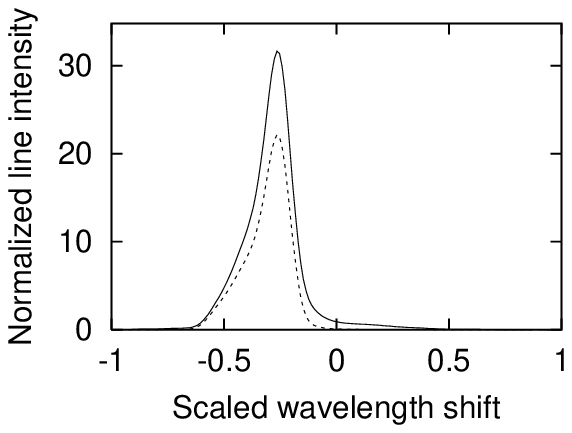} 
	& \includegraphics[width=5cm]{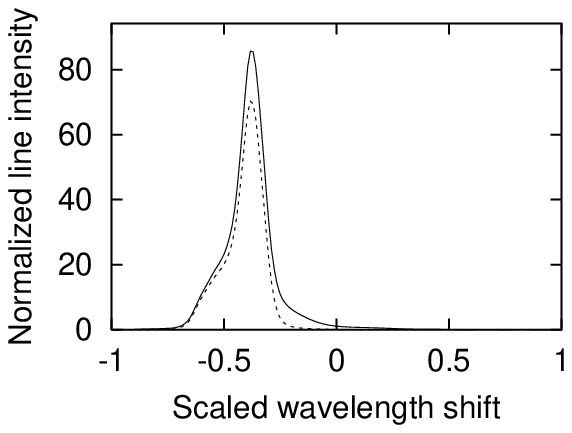} \\
\raisebox{1.5cm}{$\theta = 45\degr$}
	& \includegraphics[width=5cm]{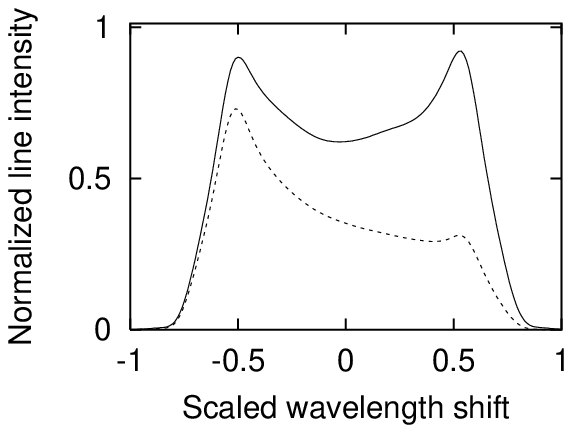} 
	& \includegraphics[width=5cm]{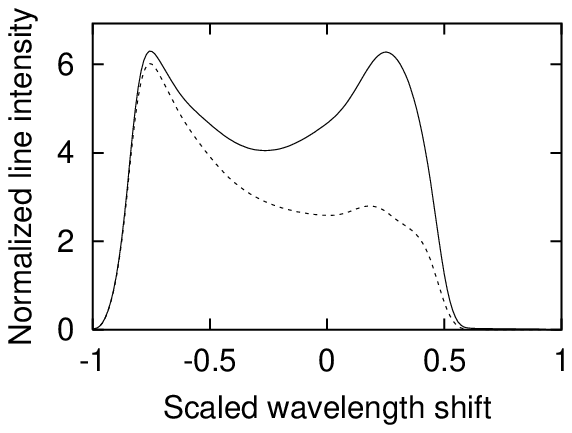} 
	& \includegraphics[width=5cm]{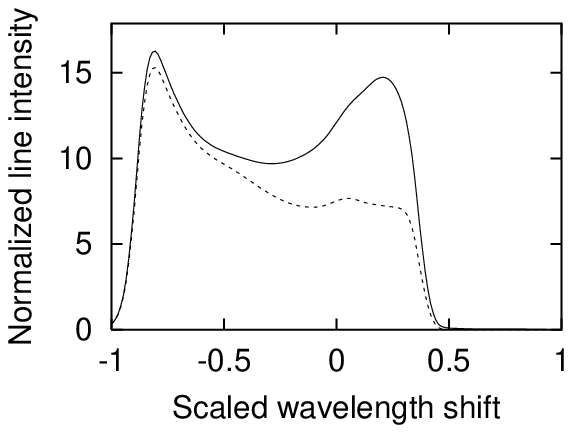} \\
\raisebox{1.5cm}{$\theta = 90\degr$}
	& \includegraphics[width=5cm]{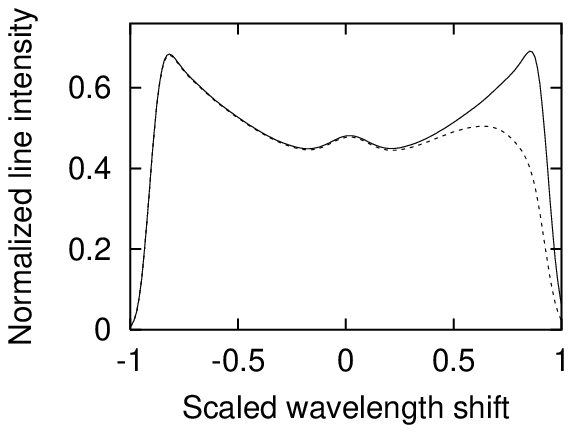} 
	& \includegraphics[width=5cm]{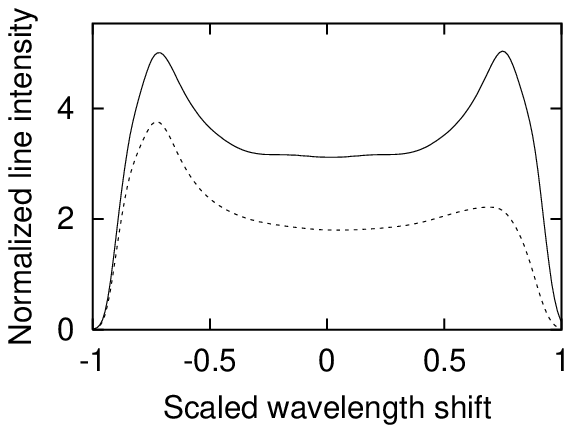} 
	& \includegraphics[width=5cm]{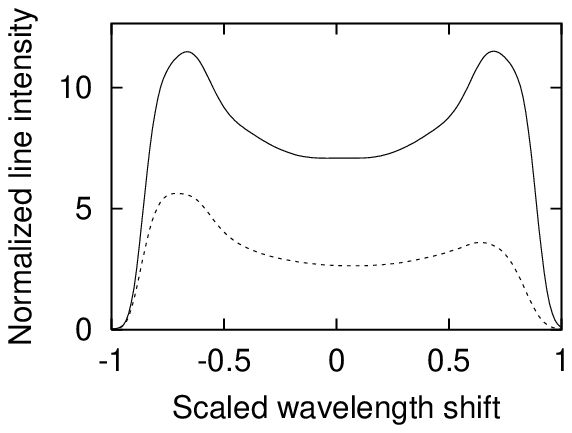} \\
\raisebox{1.5cm}{$\theta = 135\degr$}
	& \includegraphics[width=5cm]{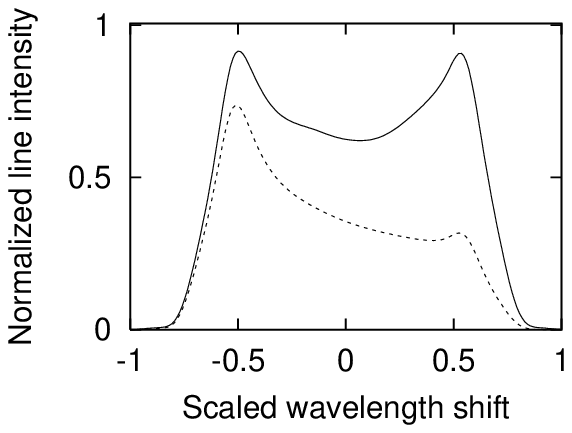} 
	& \includegraphics[width=5cm]{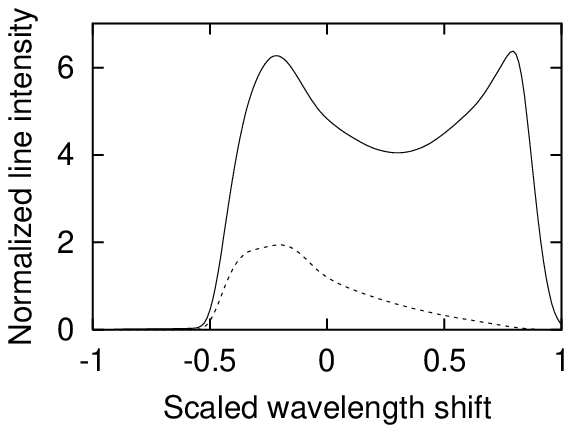} 
	& \includegraphics[width=5cm]{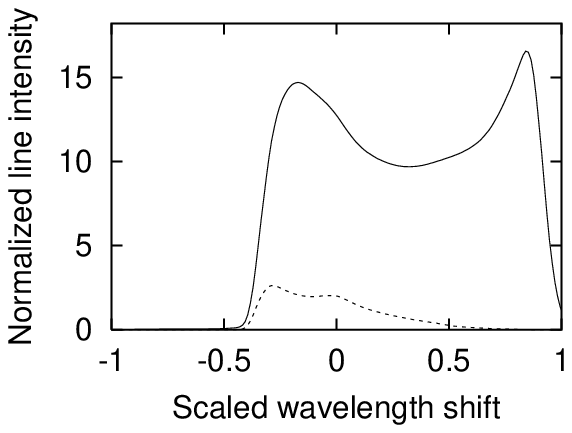} \\
\raisebox{1.5cm}{$\theta = 180\degr$}
	& \includegraphics[width=5cm]{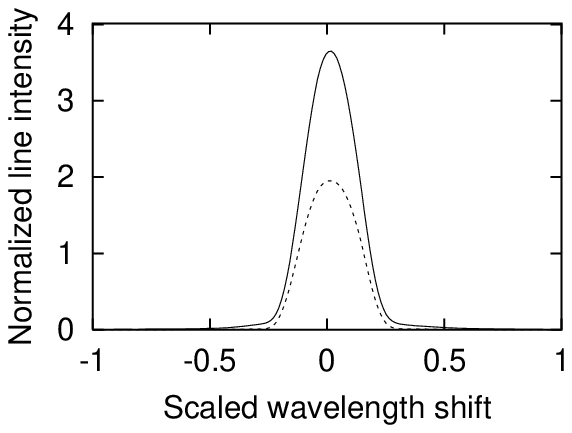} 
	& \includegraphics[width=5cm]{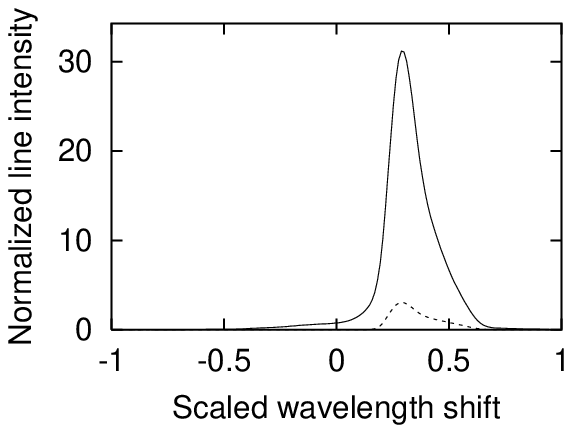} 
	& \includegraphics[width=5cm]{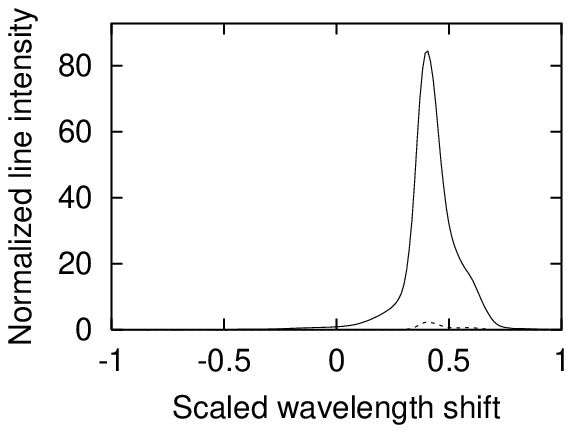} \\
\end{tabular}
\caption{Unabsorbed (solid) and absorbed (dashed) \OVIII\ \Lyalpha\ line profiles for a range of wind momentum ratios
\R\ and viewing angles $\theta$. The profiles are for systems with $\MdotA = 1,5,9 \times 10^{-6} \Msolpy$,
$\MdotB = 1 \times 10^{-6} \Msolpy$ and $\vA = \vB = 2000 \kmps$.}
\label{fig:o8lineprofiles}
\end{figure*}

\begin{figure*}
\begin{tabular}{cccc}
& $\R = 1$ & $\R = \sqrt{5}$ & $\R = 3$ \\
\raisebox{1.5cm}{$\theta = 0\degr$}
	& \includegraphics[width=5cm]{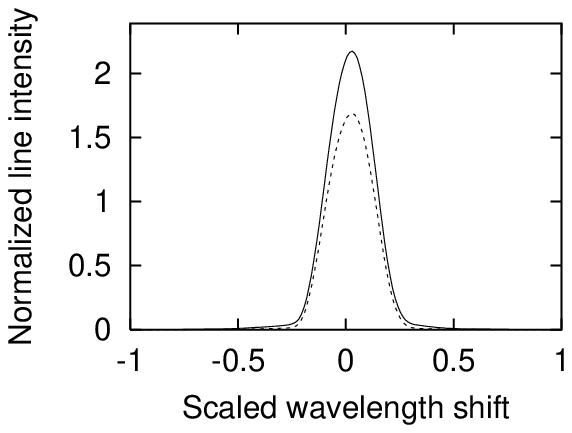} 
	& \includegraphics[width=5cm]{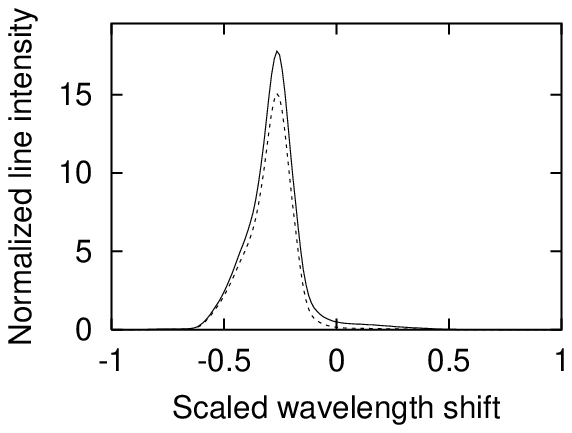} 
	& \includegraphics[width=5cm]{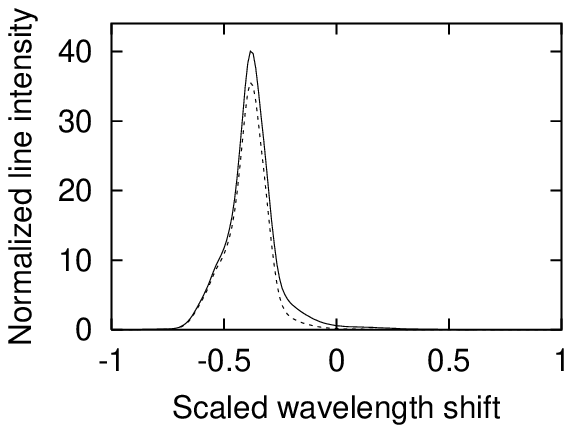} \\
\raisebox{1.5cm}{$\theta = 45\degr$}
	& \includegraphics[width=5cm]{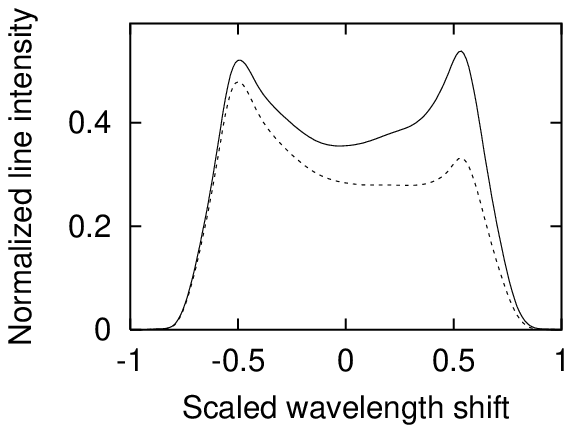} 
	& \includegraphics[width=5cm]{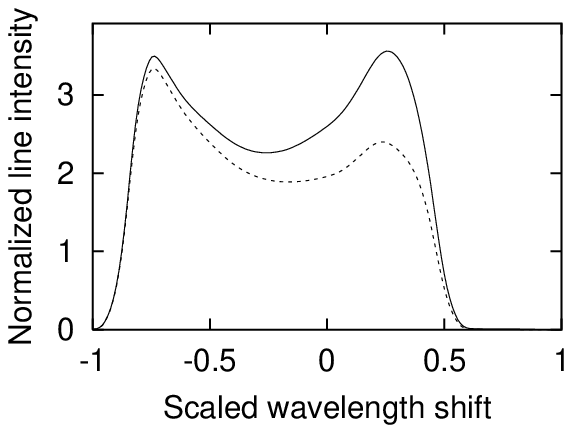} 
	& \includegraphics[width=5cm]{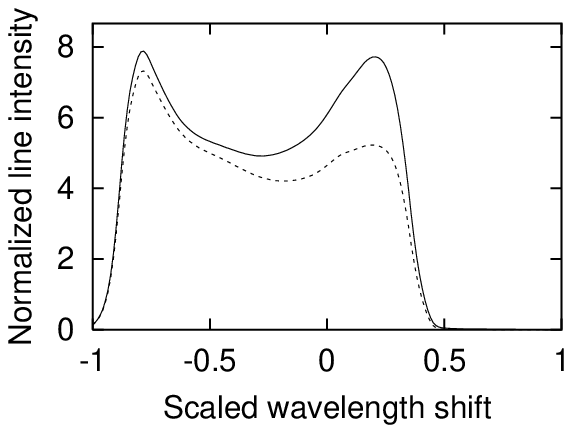} \\
\raisebox{1.5cm}{$\theta = 90\degr$}
	& \includegraphics[width=5cm]{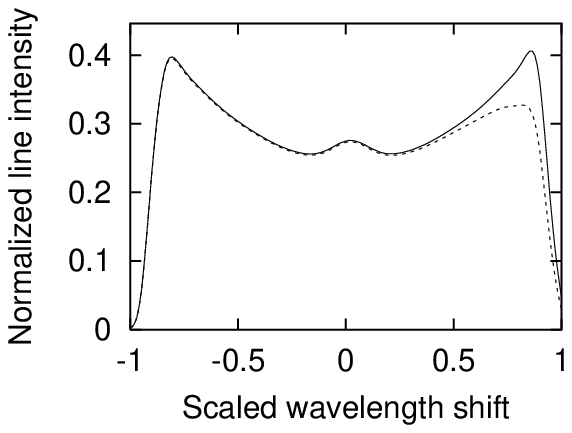} 
	& \includegraphics[width=5cm]{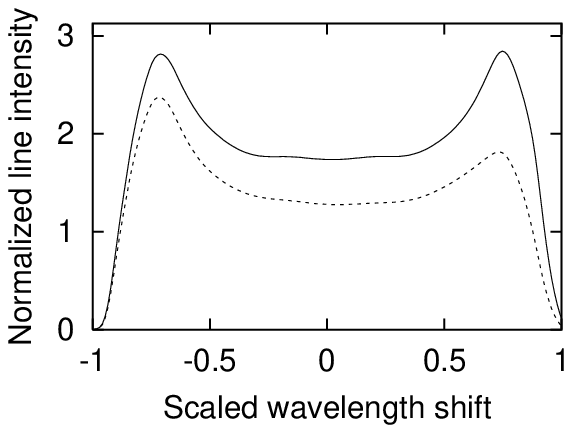} 
	& \includegraphics[width=5cm]{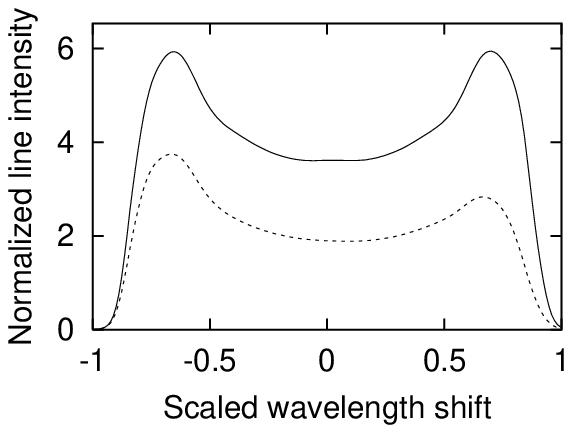} \\
\raisebox{1.5cm}{$\theta = 135\degr$}
	& \includegraphics[width=5cm]{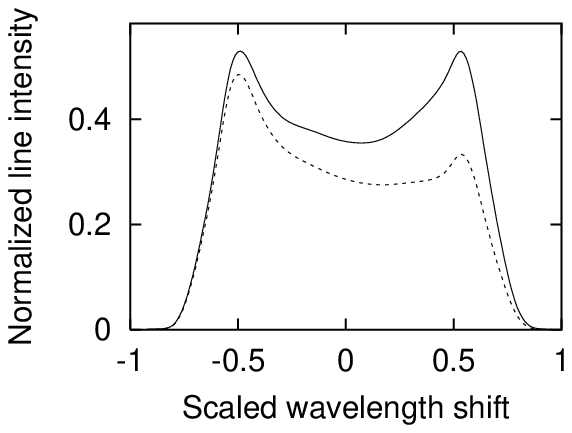} 
	& \includegraphics[width=5cm]{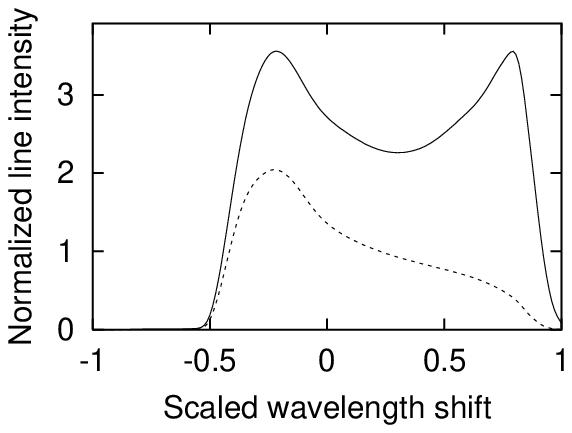} 
	& \includegraphics[width=5cm]{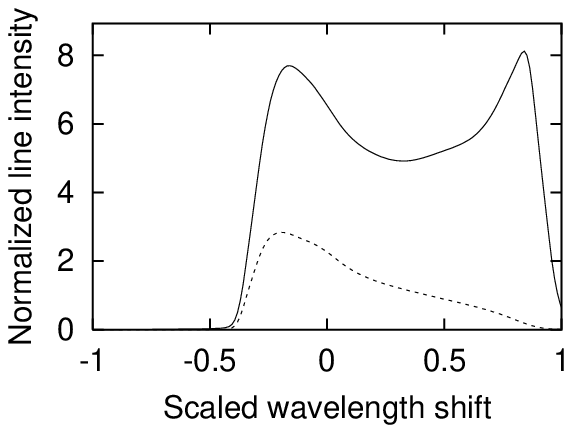} \\
\raisebox{1.5cm}{$\theta = 180\degr$}
	& \includegraphics[width=5cm]{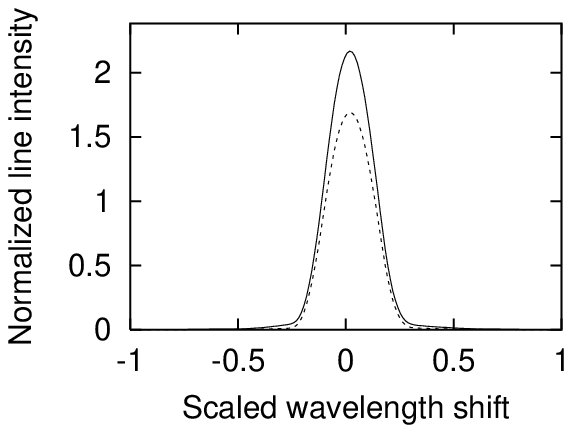} 
	& \includegraphics[width=5cm]{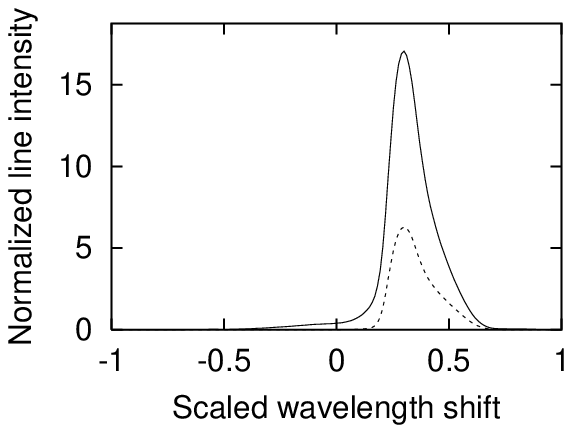} 
	& \includegraphics[width=5cm]{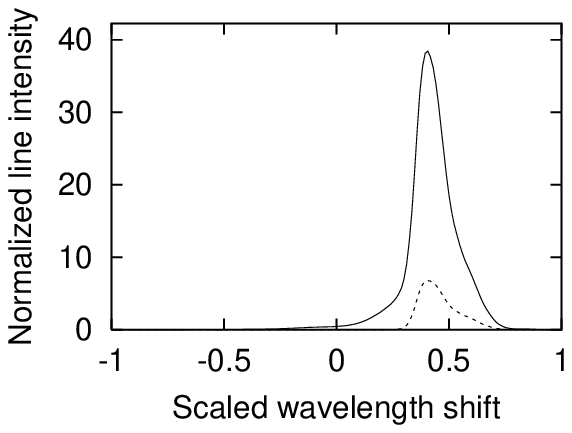} \\
\end{tabular}
\caption{Unabsorbed (solid) and absorbed (dashed) \NeX\ \Lyalpha\ line profiles for a range of wind momentum ratios
\R\ and viewing angles $\theta$. The profiles are for systems with $\MdotA = 1,5,9 \times 10^{-6} \Msolpy$,
$\MdotB = 1 \times 10^{-6} \Msolpy$ and $\vA = \vB = 2000 \kmps$.}
\label{fig:ne10lineprofiles}
\end{figure*}

\begin{figure*}
\begin{tabular}{cccc}
& $\R = 1$ & $\R = \sqrt{5}$ & $\R = 3$ \\
\raisebox{1.5cm}{$\theta = 0\degr$}
	& \includegraphics[width=5cm]{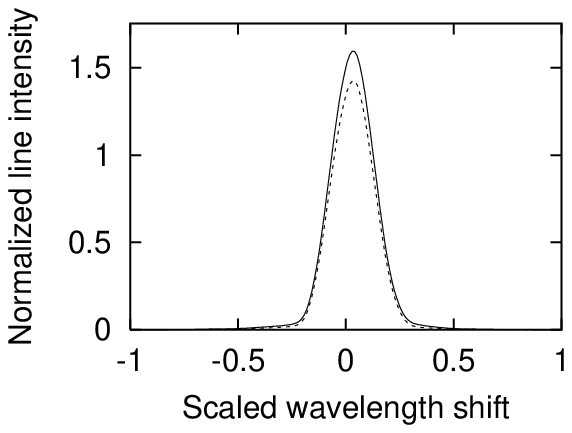} 
	& \includegraphics[width=5cm]{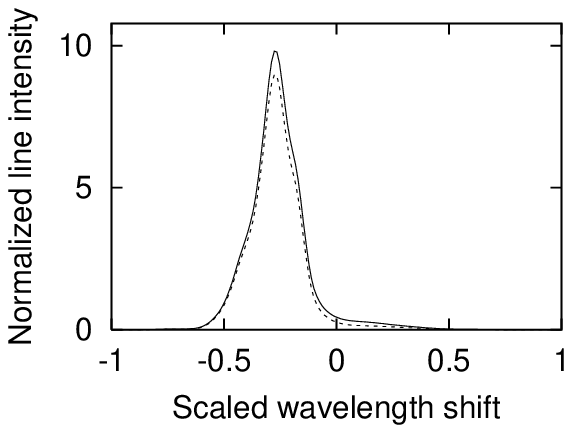} 
	& \includegraphics[width=5cm]{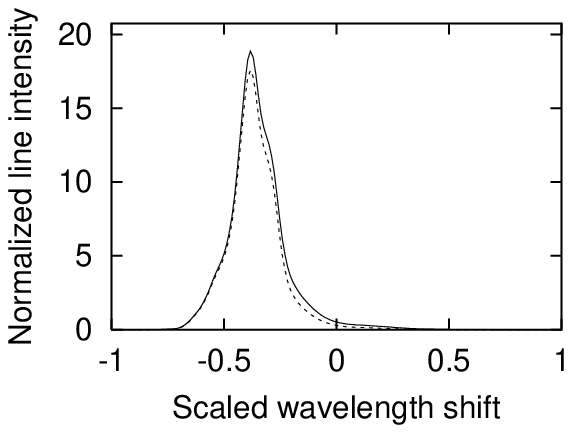} \\
\raisebox{1.5cm}{$\theta = 45\degr$}
	& \includegraphics[width=5cm]{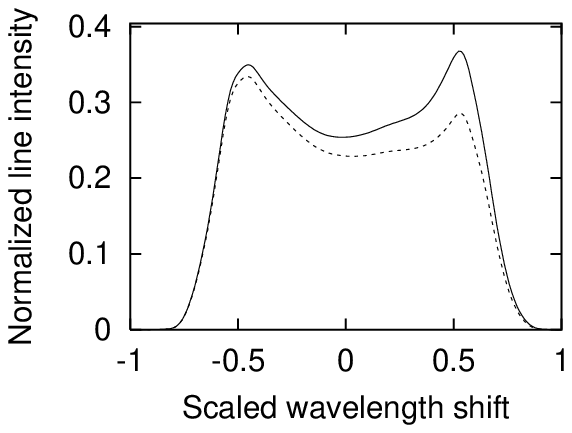} 
	& \includegraphics[width=5cm]{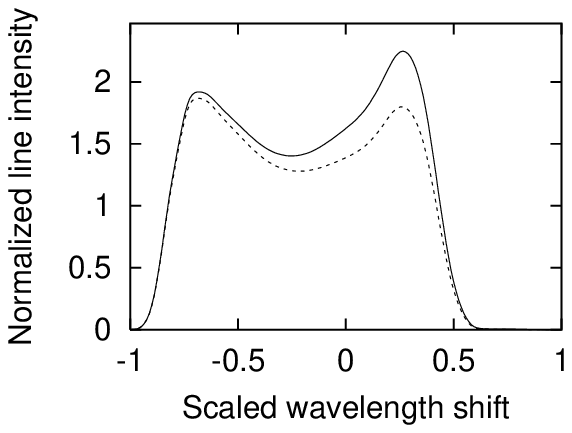} 
	& \includegraphics[width=5cm]{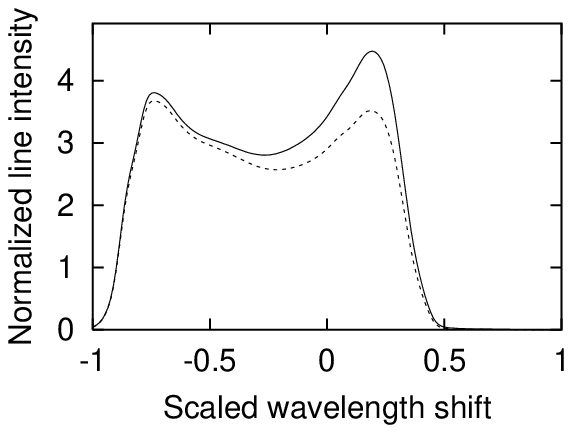} \\
\raisebox{1.5cm}{$\theta = 90\degr$}
	& \includegraphics[width=5cm]{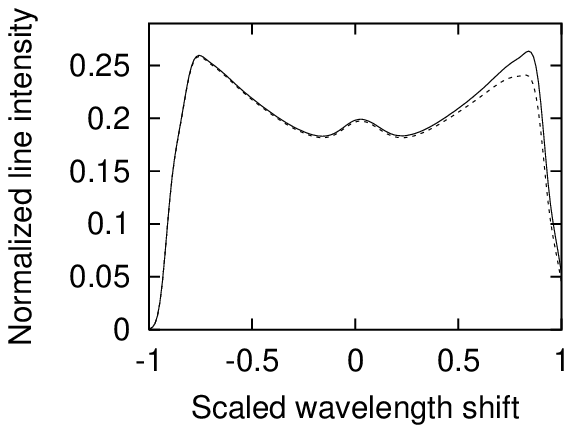} 
	& \includegraphics[width=5cm]{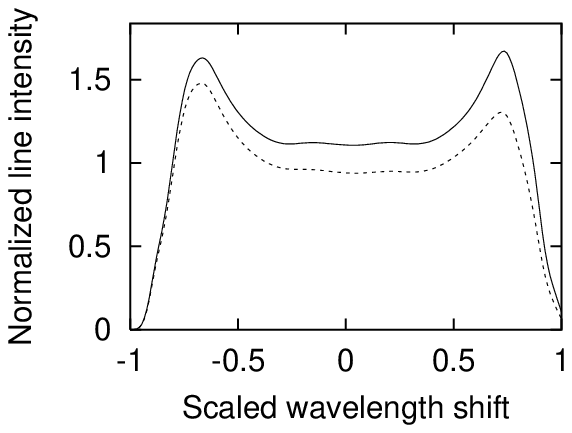} 
	& \includegraphics[width=5cm]{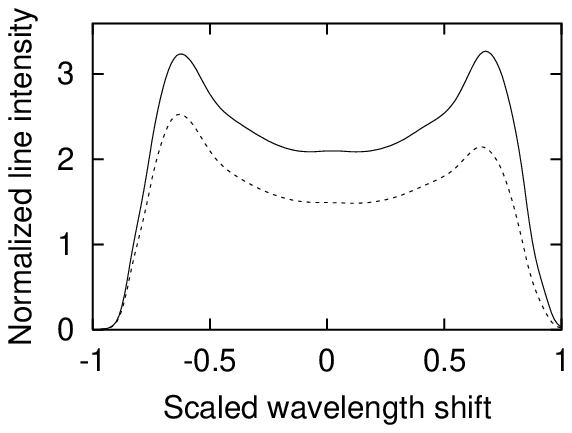} \\
\raisebox{1.5cm}{$\theta = 135\degr$}
	& \includegraphics[width=5cm]{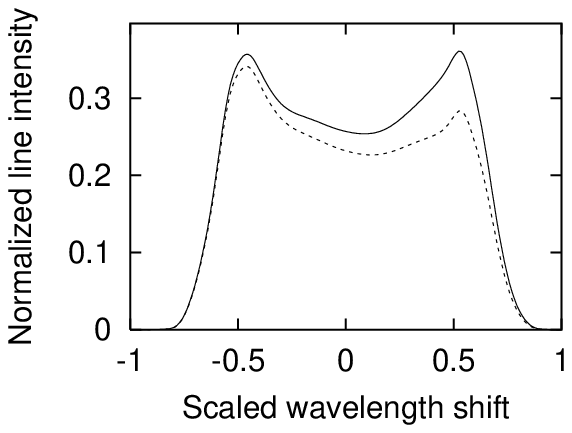} 
	& \includegraphics[width=5cm]{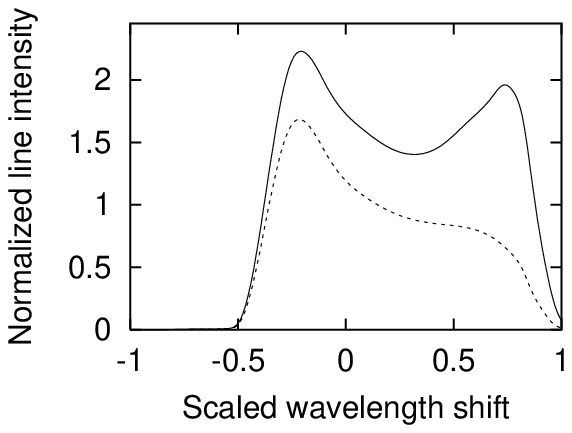} 
	& \includegraphics[width=5cm]{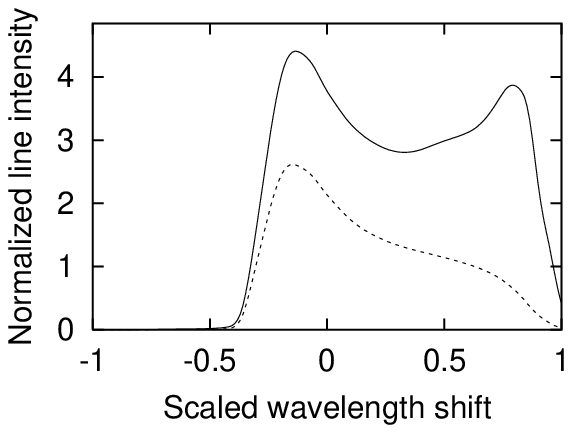} \\
\raisebox{1.5cm}{$\theta = 180\degr$}
	& \includegraphics[width=5cm]{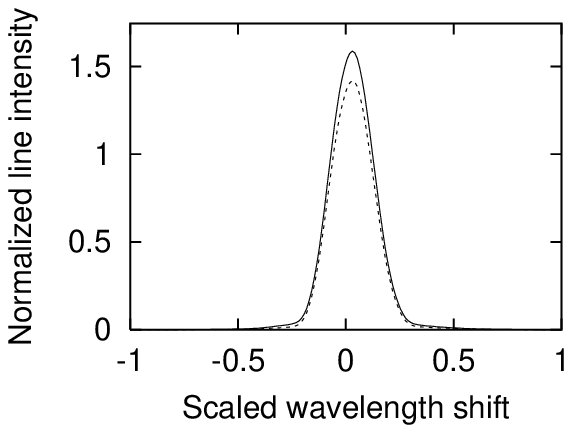} 
	& \includegraphics[width=5cm]{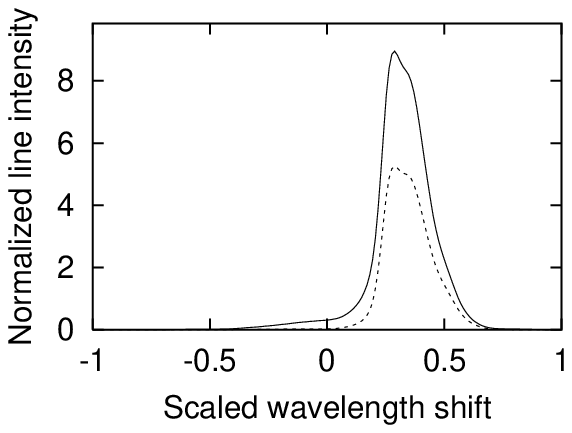} 
	& \includegraphics[width=5cm]{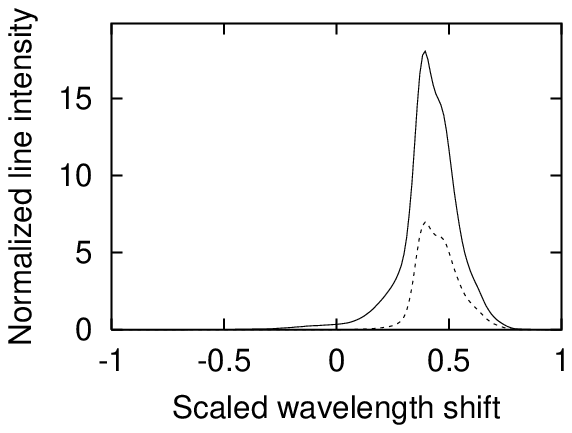} \\
\end{tabular}
\caption{Unabsorbed (solid) and absorbed (dashed) \MgXII\ \Lyalpha\ line profiles for a range of wind momentum ratios
\R\ and viewing angles $\theta$. The profiles are for systems with $\MdotA = 1,5,9 \times 10^{-6} \Msolpy$,
$\MdotB = 1 \times 10^{-6} \Msolpy$ and $\vA = \vB = 2000 \kmps$.}
\label{fig:mg12lineprofiles}
\end{figure*}

\begin{figure*}
\begin{tabular}{cccc}
& $\R = 1$ & $\R = \sqrt{5}$ & $\R = 3$ \\
\raisebox{1.5cm}{$\theta = 0\degr$}
	& \includegraphics[width=5cm]{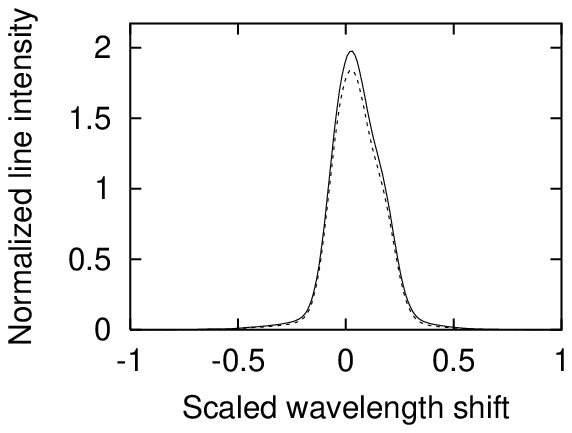} 
	& \includegraphics[width=5cm]{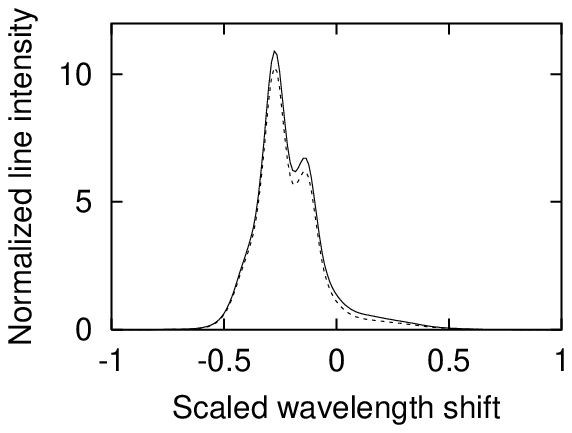} 
	& \includegraphics[width=5cm]{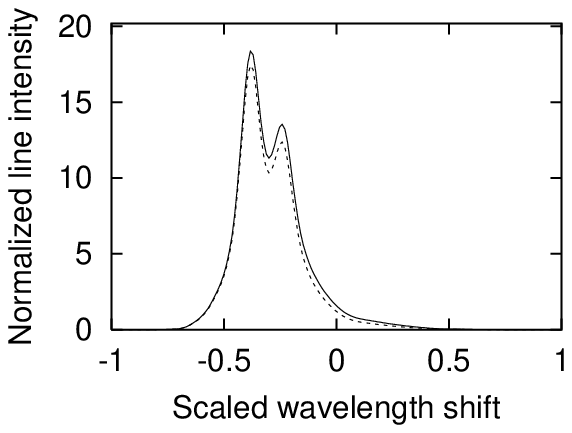} \\
\raisebox{1.5cm}{$\theta = 45\degr$}
	& \includegraphics[width=5cm]{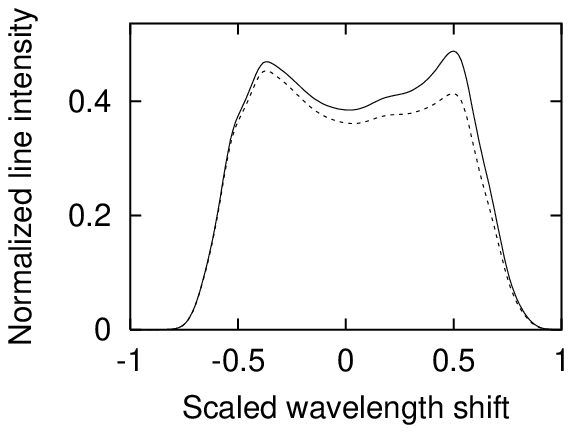} 
	& \includegraphics[width=5cm]{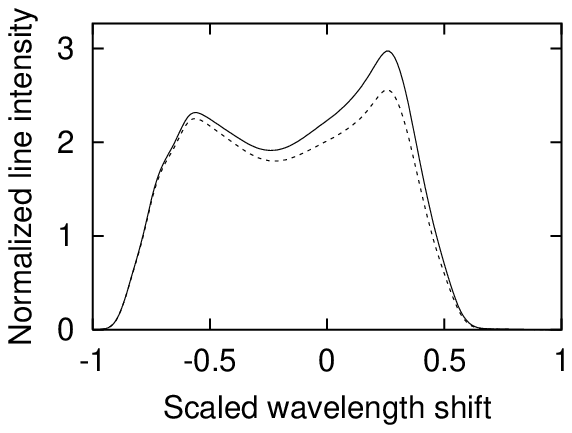} 
	& \includegraphics[width=5cm]{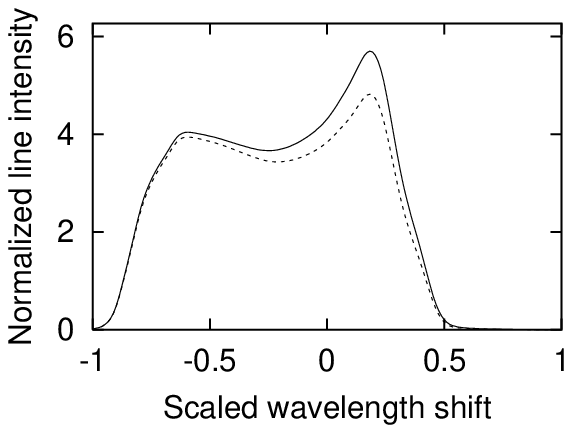} \\
\raisebox{1.5cm}{$\theta = 90\degr$}
	& \includegraphics[width=5cm]{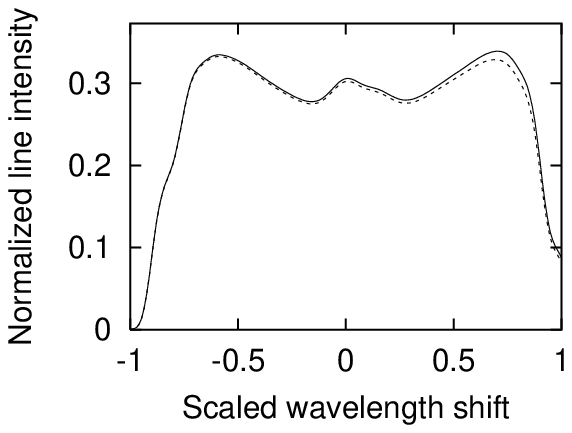} 
	& \includegraphics[width=5cm]{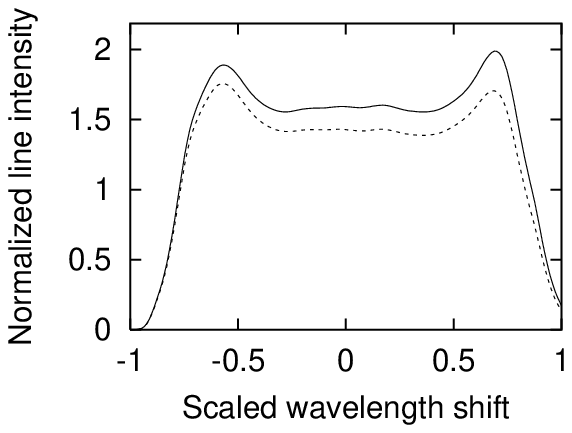} 
	& \includegraphics[width=5cm]{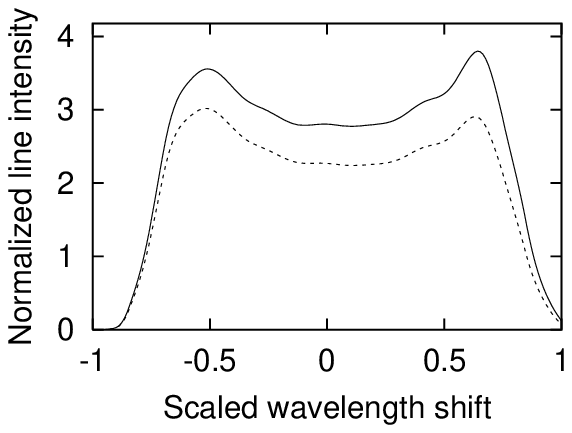} \\
\raisebox{1.5cm}{$\theta = 135\degr$}
	& \includegraphics[width=5cm]{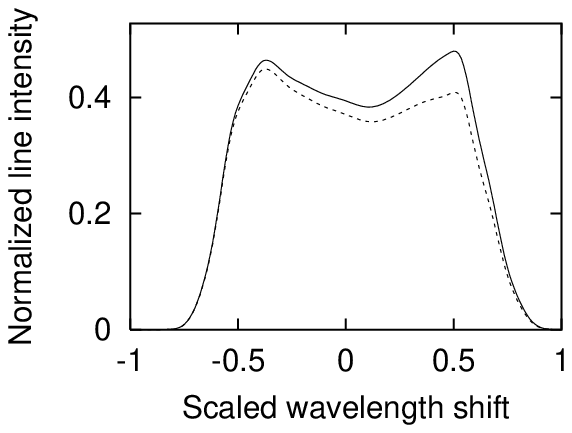} 
	& \includegraphics[width=5cm]{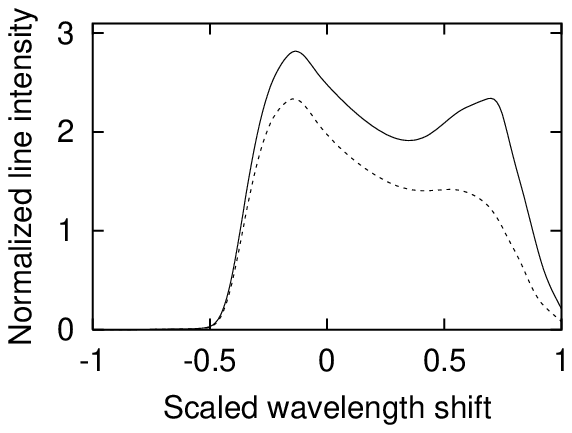} 
	& \includegraphics[width=5cm]{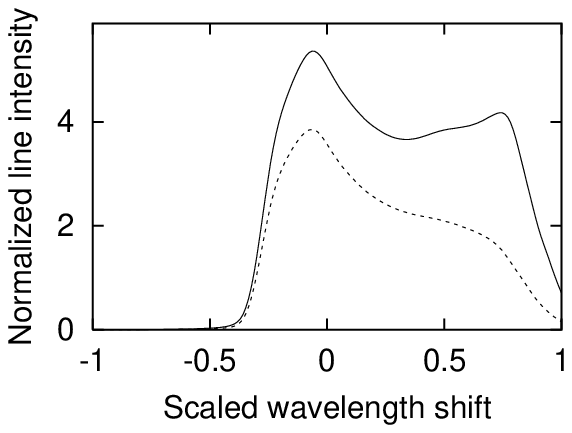} \\
\raisebox{1.5cm}{$\theta = 180\degr$}
	& \includegraphics[width=5cm]{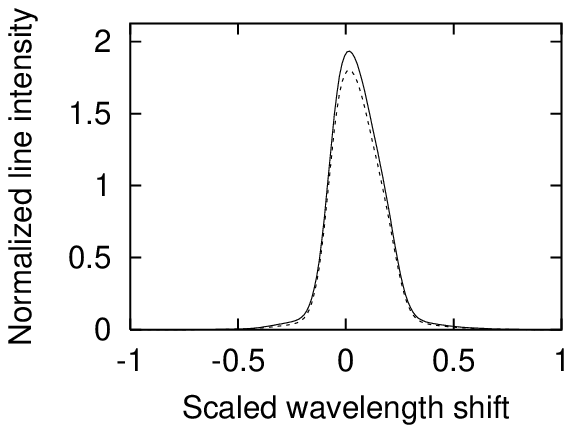} 
	& \includegraphics[width=5cm]{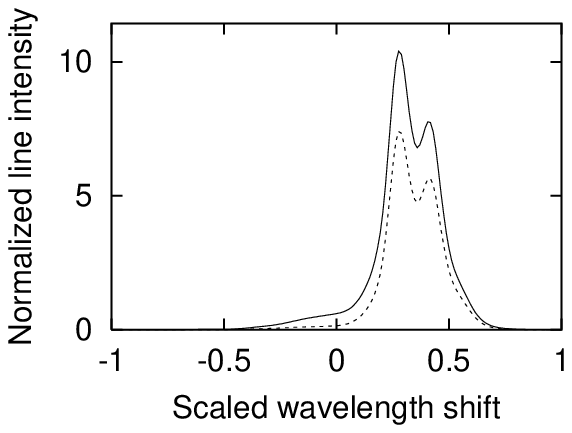} 
	& \includegraphics[width=5cm]{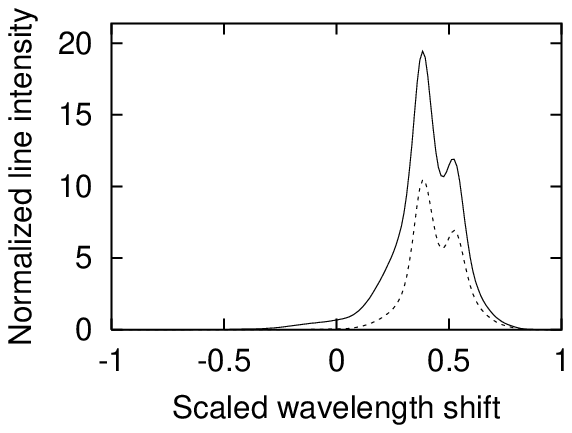} \\
\end{tabular}
\caption{Unabsorbed (solid) and absorbed (dashed) \SiXIV\ \Lyalpha\ line profiles for a range of wind momentum ratios
\R\ and viewing angles $\theta$. The profiles are for systems with $\MdotA = 1,5,9 \times 10^{-6} \Msolpy$,
$\MdotB = 1 \times 10^{-6} \Msolpy$ and $\vA = \vB = 2000 \kmps$.}
\label{fig:si14lineprofiles}
\end{figure*}

\begin{figure*}
\begin{tabular}{cccc}
& $\R = 1$ & $\R = \sqrt{5}$ & $\R = 3$ \\
\raisebox{1.5cm}{$\theta = 0\degr$}
	& \includegraphics[width=5cm]{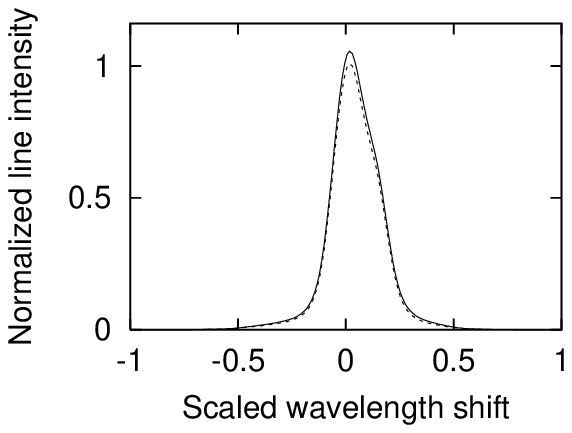} 
	& \includegraphics[width=5cm]{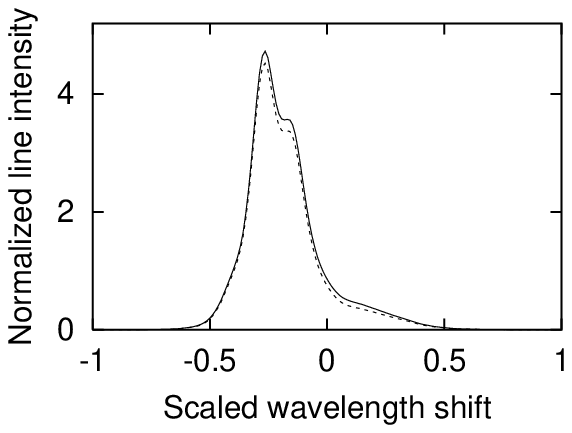} 
	& \includegraphics[width=5cm]{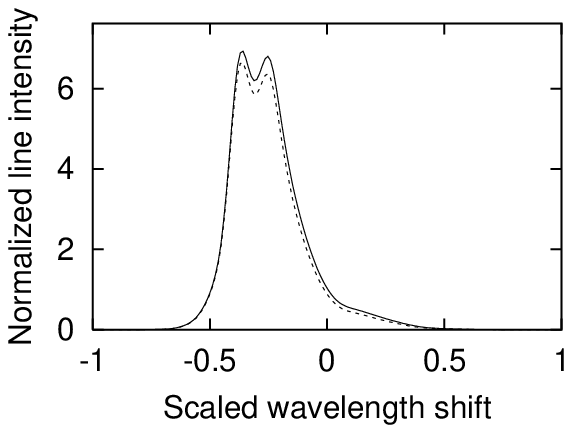} \\
\raisebox{1.5cm}{$\theta = 45\degr$}
	& \includegraphics[width=5cm]{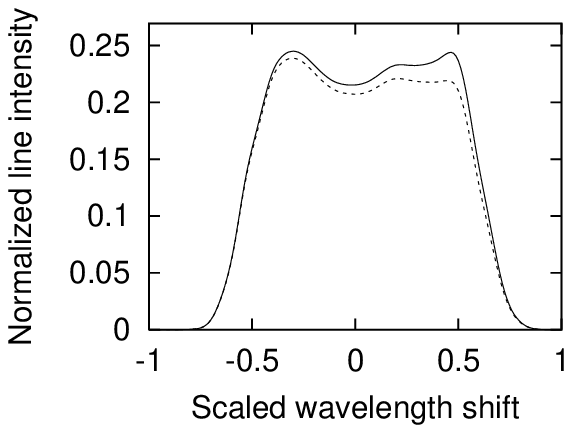} 
	& \includegraphics[width=5cm]{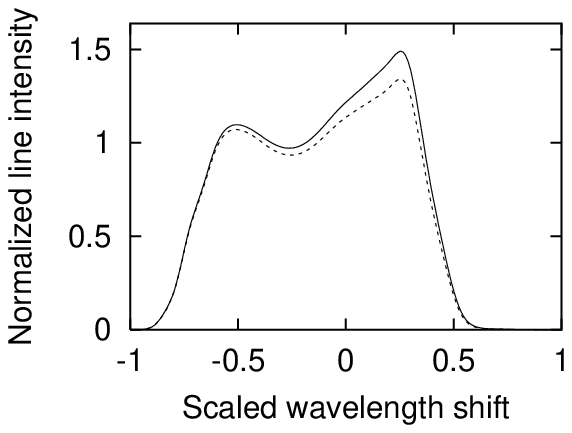} 
	& \includegraphics[width=5cm]{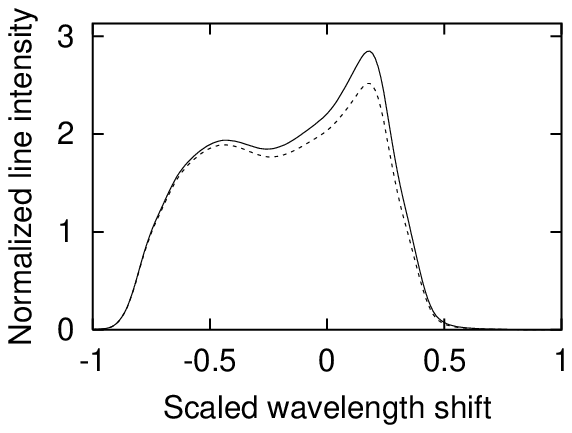} \\
\raisebox{1.5cm}{$\theta = 90\degr$}
	& \includegraphics[width=5cm]{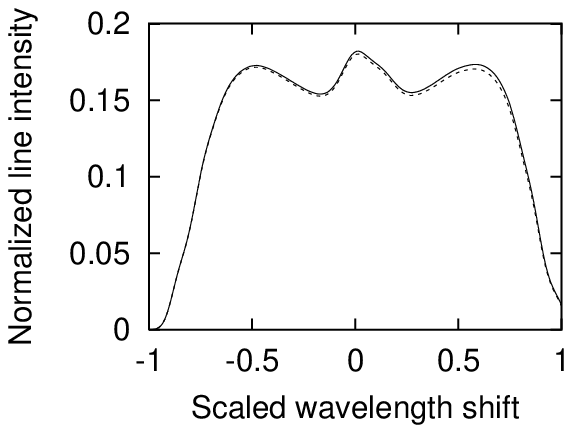} 
	& \includegraphics[width=5cm]{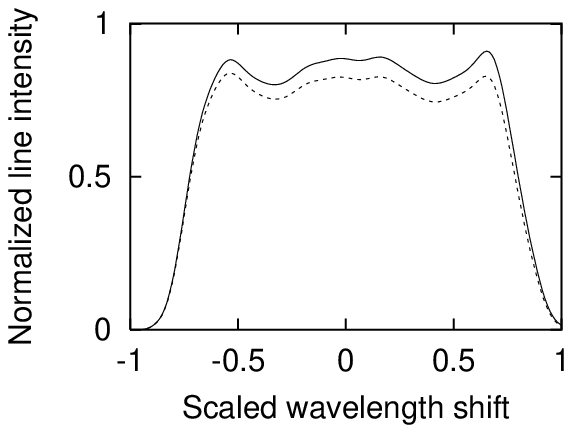} 
	& \includegraphics[width=5cm]{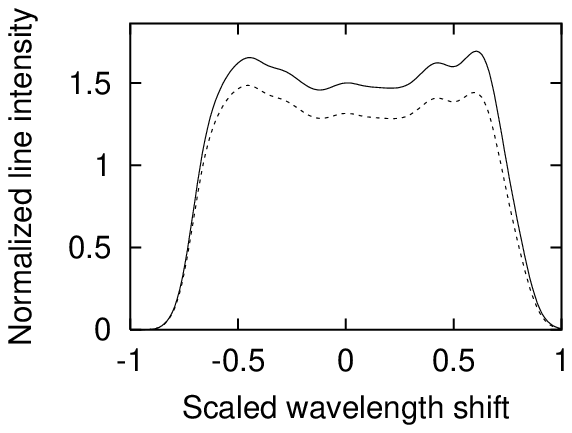} \\
\raisebox{1.5cm}{$\theta = 135\degr$}
	& \includegraphics[width=5cm]{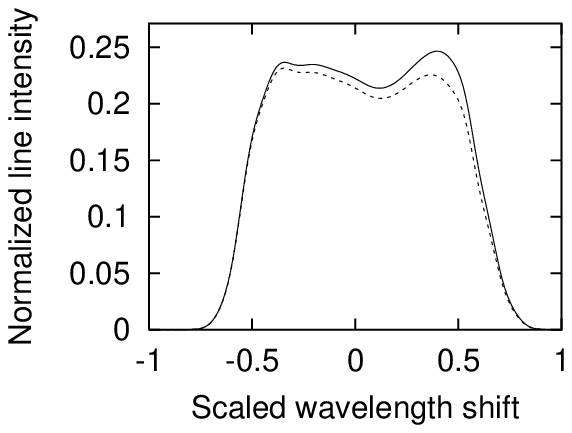} 
	& \includegraphics[width=5cm]{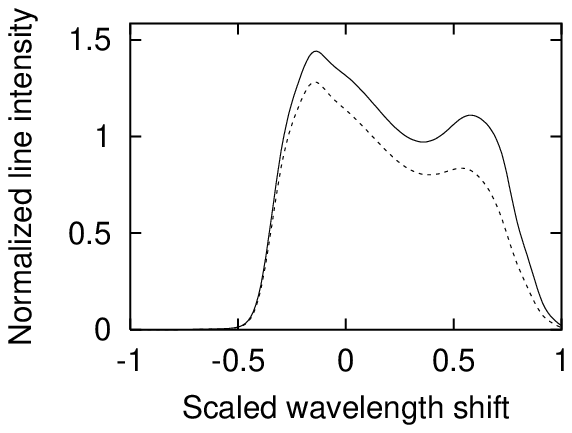} 
	& \includegraphics[width=5cm]{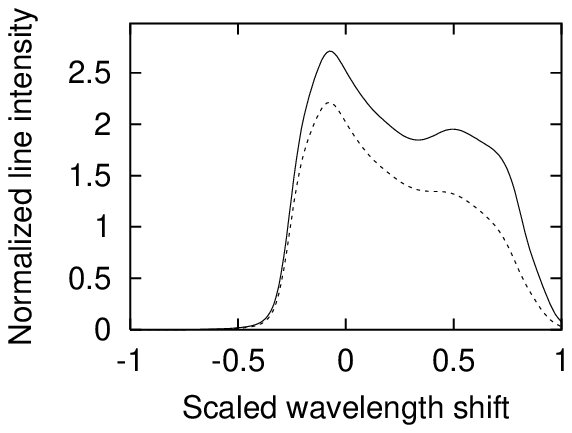} \\
\raisebox{1.5cm}{$\theta = 180\degr$}
	& \includegraphics[width=5cm]{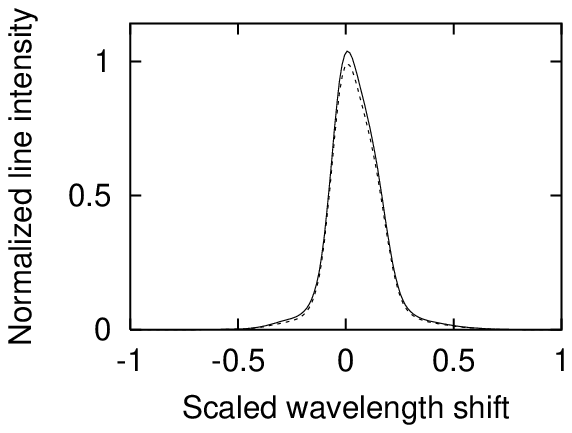} 
	& \includegraphics[width=5cm]{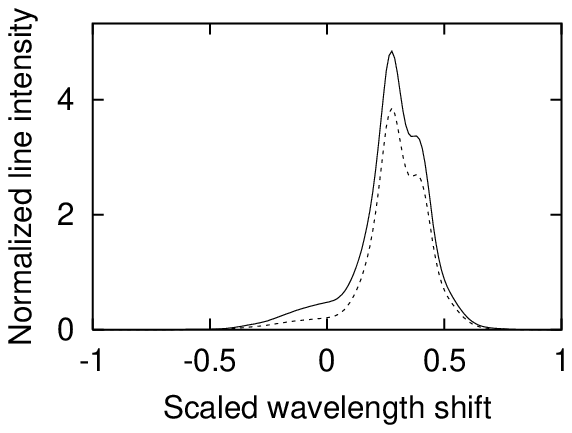} 
	& \includegraphics[width=5cm]{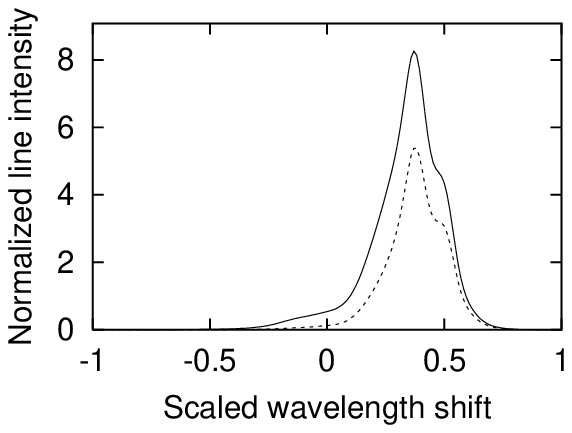} \\
\end{tabular}
\caption{Unabsorbed (solid) and absorbed (dashed) \SXVI\ \Lyalpha\ line profiles for a range of wind momentum ratios
\R\ and viewing angles $\theta$. The profiles are for systems with $\MdotA = 1,5,9 \times 10^{-6} \Msolpy$,
$\MdotB = 1 \times 10^{-6} \Msolpy$ and $\vA = \vB = 2000 \kmps$.}
\label{fig:s16lineprofiles}
\end{figure*}

%%%%%%%%%%%%%%%%%%%%%%%%%%%%%%%%%%%%%%%%%%%%%%%%%%%%%%%%%%%%%%%%%%%%%%%%%%%%%%%%

\subsection{The profiles}

Figs.~\ref{fig:o8lineprofiles} to \ref{fig:s16lineprofiles} show theoretical \Lyalpha\ line profiles for \OVIII,
\NeX, \MgXII, \SiXIV\ and \SXVI, respectively. The profiles have been calculated from single snapshots from
the hydrodynamical simulations. The model CWBs have
$\MdotB = 1 \times 10^{-6} \Msolpy$ and $\vA = \vB = 2000 \kmps$; the wind momentum ratio \R\ is adjusted
by varying \MdotA. In all cases, profiles are shown for 3 values of \R\ and for 5 different viewing angles 
$\theta$, where $\theta$ is the angle between the line-of-sight and
the line-of-centres, with $\theta = 0\degr$ corresponding to viewing the system from behind the secondary star.
For a circular orbit, $\theta$ is related to the inclination $i$ and the orbital phase angle $\phi$ by

\begin{equation}
\cos \theta = \sin i \cos \phi
\label{eq:ViewingAngle}
\end{equation}

\noindent
where $\phi = 0\degr$ corresponds to the secondary being in front.
Profiles with and without absorption by the stellar winds are shown.
The wavelength shifts are calculated relative to the rest wavelength of the brighter component of the \Lyalpha\ lines,
and have been scaled to the wind velocity. The unabsorbed \OVIII\ \Lyalpha\ line for $\R = 1$ has been normalized
such that its total luminosity is equal to unity. The other lines are normalized relative to this line,
preserving the relative line luminosities.

The secondary components of the \Lyalpha\ lines are only noticeable for the shorter wavelength lines. This is because
the wavelength difference between the components (0.004--0.006\angstrom; see Table~\ref{table:LineWavelengths})
corresponds to a larger velocity shift for these lines. In practice, after the lines have been folded with the
instrumental response of the \chandra\ or \xmm\ gratings, the two components would not be resolvable. However,
including both components of the \Lyalpha\ lines gives broader lines than one would get by considering only
the brighter component.

%%%%%%%%%%%%%%%%%%%%%%%%%%%%%%%%%%%%%%%%%%%%%%%%%%%%%%%%%%%%%%%%%%%%%%%%%%%%%%%%%%%%%%%%%%%%%%%%%%%%%%%%%%%%%%%%%%%%%%%%

\subsection{Variation of line parameters with viewing angle and wind momentum ratio}
\label{subsec:VariationWithThetaAndR}

\begin{figure}
  \includegraphics[width=8cm]{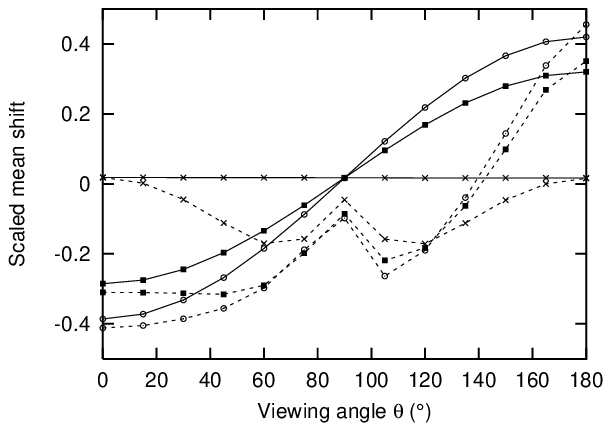}
  \includegraphics[width=8cm]{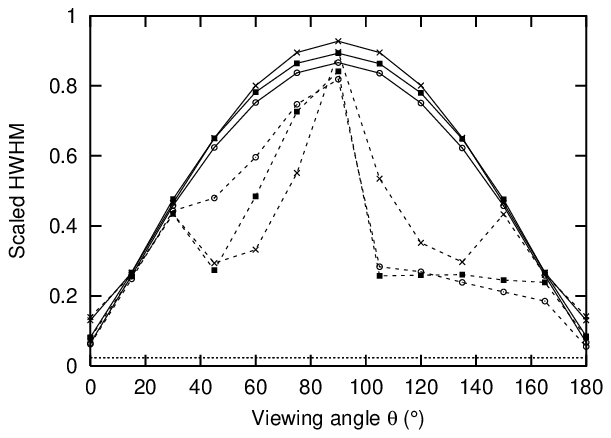}
  \includegraphics[width=8cm]{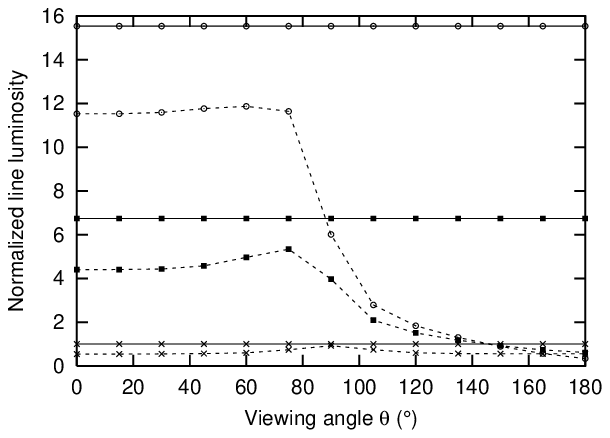}
  \caption{Graphs showing how the mean wavelength shift, HWHM, and normalized line
  luminosity of the unabsorbed (solid) and absorbed (dashed) \OVIII\ \Lyalpha\ line
  profile vary with viewing angle. Results for $\R = 1$, $\sqrt{5}$ and 3 are represented
  by crosses, solid squares and open circles, respectively.}
  \label{fig:o8linestats}
\end{figure}

\begin{figure}
  \includegraphics[width=8cm]{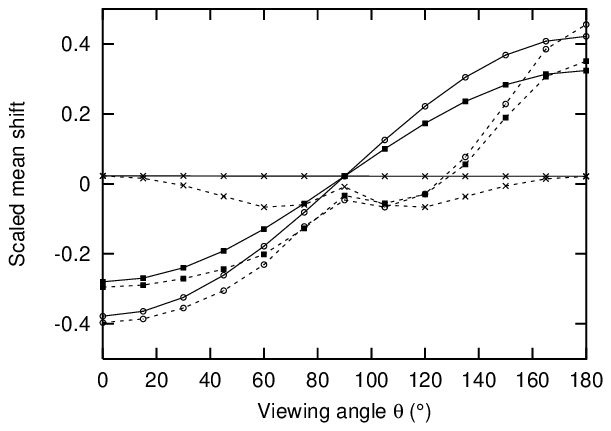}
  \includegraphics[width=8cm]{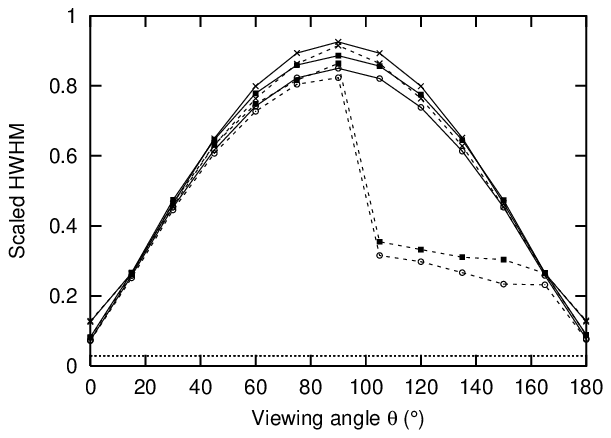}
  \includegraphics[width=8cm]{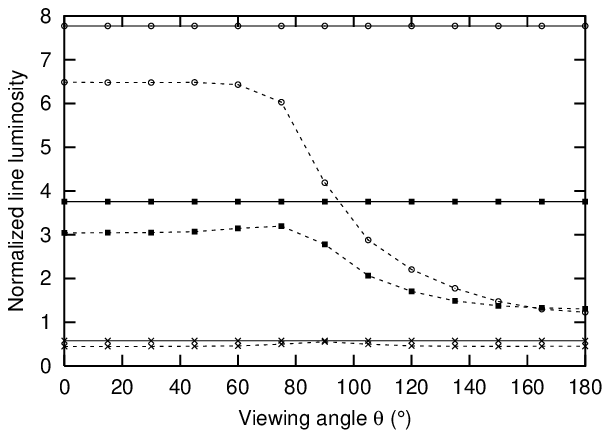}
  \caption{Graphs showing how the mean wavelength shift, HWHM, and normalized line
  luminosity of the unabsorbed (solid) and absorbed (dashed) \NeX\ \Lyalpha\ line
  profile vary with viewing angle. Results for $\R = 1$, $\sqrt{5}$ and 3 are represented
  by crosses, solid squares and open circles, respectively.}
  \label{fig:ne10linestats}
\end{figure}

\begin{figure}
  \includegraphics[width=8cm]{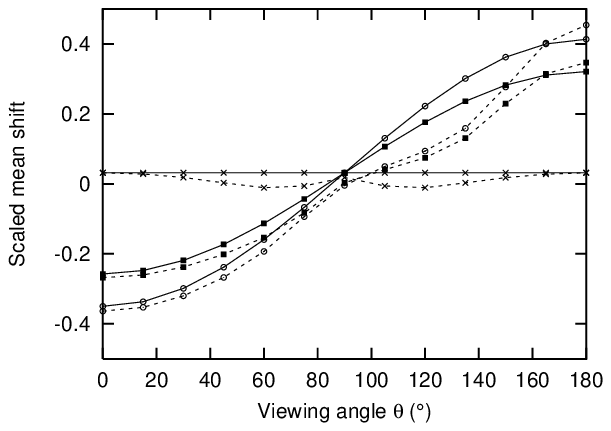}
  \includegraphics[width=8cm]{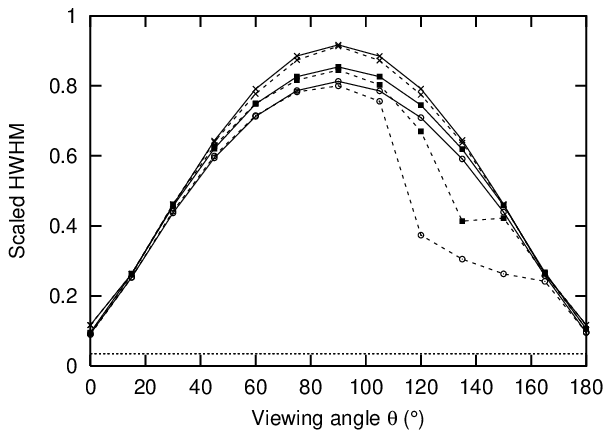}
  \includegraphics[width=8cm]{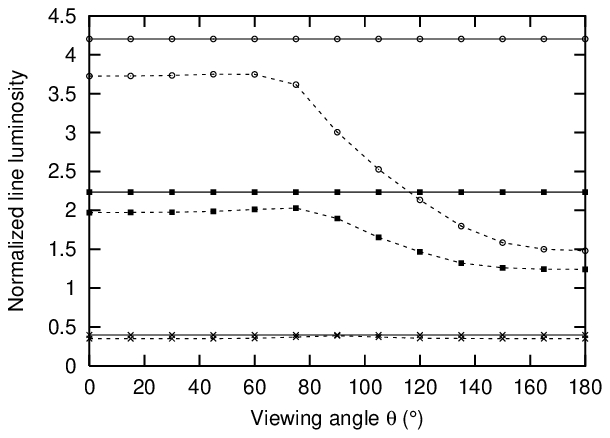}
  \caption{Graphs showing how the mean wavelength shift, HWHM, and normalized line
  luminosity of the unabsorbed (solid) and absorbed (dashed) \MgXII\ \Lyalpha\ line
  profile vary with viewing angle. Results for $\R = 1$, $\sqrt{5}$ and 3 are represented
  by crosses, solid squares and open circles, respectively.}
  \label{fig:mg12linestats}
\end{figure}

\begin{figure}
  \includegraphics[width=8cm]{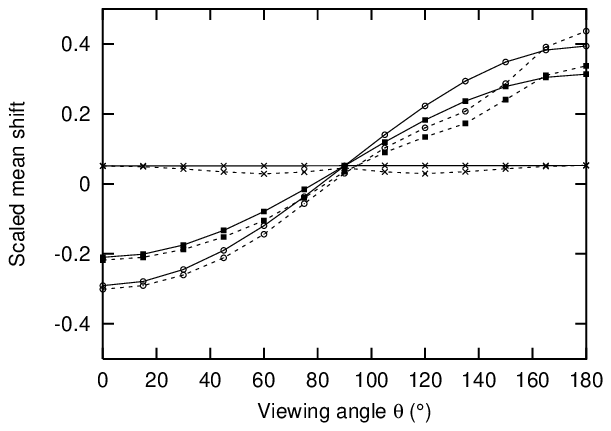}
  \includegraphics[width=8cm]{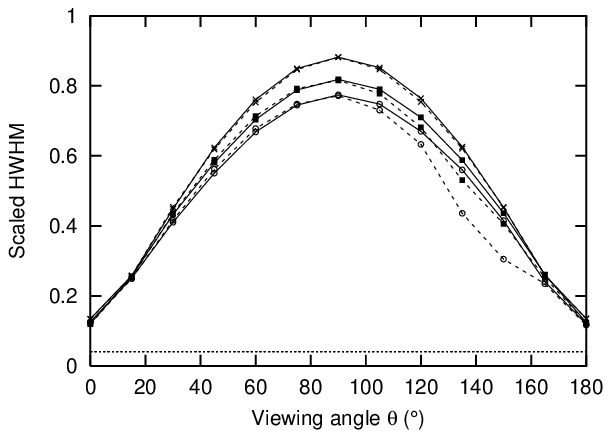}
  \includegraphics[width=8cm]{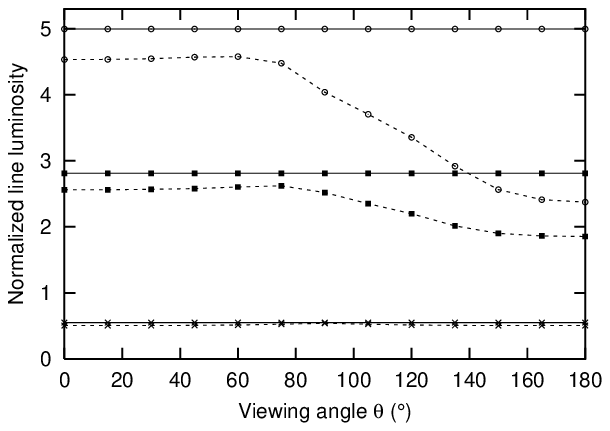}
  \caption{Graphs showing how the mean wavelength shift, HWHM, and normalized line
  luminosity of the unabsorbed (solid) and absorbed (dashed) \SiXIV\ \Lyalpha\ line
  profile vary with viewing angle. Results for $\R = 1$, $\sqrt{5}$ and 3 are represented
  by crosses, solid squares and open circles, respectively.}
  \label{fig:si14linestats}
\end{figure}

\begin{figure}
  \includegraphics[width=8cm]{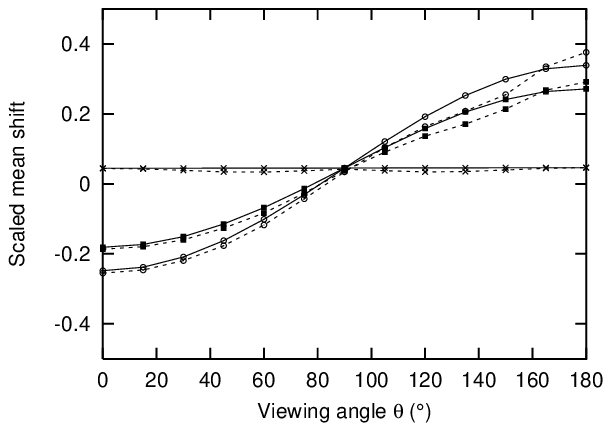}
  \includegraphics[width=8cm]{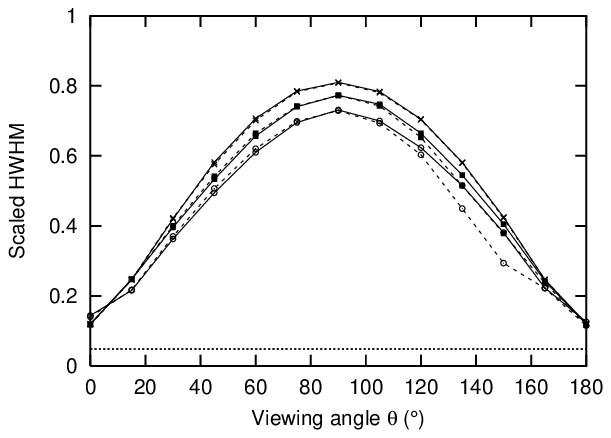}
  \includegraphics[width=8cm]{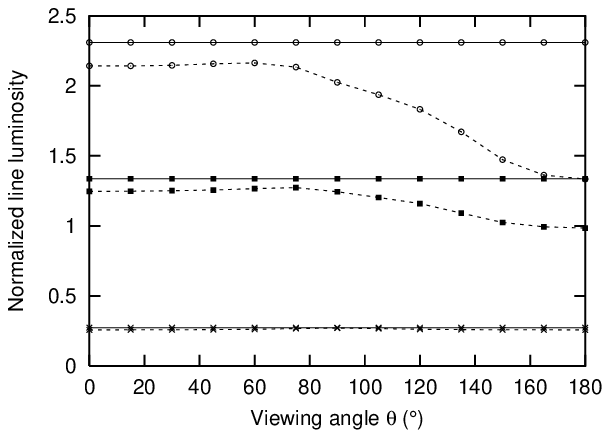}
  \caption{Graphs showing how the mean wavelength shift, HWHM, and normalized line
  luminosity of the unabsorbed (solid) and absorbed (dashed) \SXVI\ \Lyalpha\ line
  profile vary with viewing angle. Results for $\R = 1$, $\sqrt{5}$ and 3 are represented
  by crosses, solid squares and open circles, respectively.}
  \label{fig:s16linestats}
\end{figure}

Figs.~\ref{fig:o8linestats} to \ref{fig:s16linestats} show how line parameters vary with viewing
angle for each of the five lines. Each figure shows the variation of the mean scaled wavelength shift, the 
half width at half maximum (HWHM), and the normalized line luminosity for the unabsorbed and
absorbed line profiles for three different values of \R\ (1, $\sqrt{5}$ and 3).
The horizontal dashed lines on the HWHM graphs
show the thermal Doppler widths at the temperature of maximum line emission (for comparison).

In the equal winds case ($\R = 1$), the unabsorbed lines are approximately symmetrical and unshifted
for all viewing angles. While some slight asymmetries can be seen, these are due to numerical effects.
Theoretically, the unabsorbed lines for $\R = 1$ must be purely symmetrical.
For $\R > 1$, the lines are blueshifted when the system is viewed from the side of the
secondary ($\theta < 90 \degr$) and redshifted when the system is viewed from the
side of the primary ($\theta > 90 \degr$). The maximum shifts occur when 
$\theta = 0 \degr, 180 \degr$. This occurs because the shocked region is bent towards
the secondary when $\R > 1$. When the system is being viewed along the line of centres, the general
motion of the shocked gas is either towards or away from the observer (depending on which end
of the line of centres they are looking from), with a fairly small spread in line-of-sight velocities,
resulting in quite narrow lines. However, for other viewing angles this will not be the case --
some of the gas will be moving towards the observer, some will be moving tangentially to the line of sight,
and some will be moving away from the observer. The result of this is broader lines with smaller shifts.
The extreme case is when the system is at quadrature ($\theta = 90 \degr$). The unabsorbed lines are
very broad ($\mathrm{HWHM} \sim v_\infty$), but symmetrical and unshifted, apart from the 
presence of the secondary component on the redward side. This is exactly as expected, owing to the 
cylindrical symmetry of the hydrodynamic simulation -- for every `blob' of plasma moving towards the 
observer there's a corresponding blob on the far side of the system moving away at the same speed.
For $\theta \ne 90 \degr$, the velocity shift (towards the blue or the red, depending on the viewing
orientation) increases with \R, because the shocked region is bent more towards or away from the observer.

In all cases, the HWHMs are considerably larger than the typical thermal widths, indicating that
the line broadening is due to a wide range of line-of-sight velocities in the bulk motion of the hot
gas, rather than merely being due to simple thermal Doppler broadening.

The line luminosities increase with \R, because a larger \Mdot\ leads to a denser X-ray emitting plasma.
However, for a given \R\ the unabsorbed line luminosities do not vary with $\theta$, since these 
depend on the amount of hot plasma on the hydrodynamical grid, which is independent of viewing angle.

The above discussion pertains only to the unabsorbed line profiles. From the figures one
can clearly see that absorption has a much more significant effect on the longer wavelength lines,
such as the \OVIII\ and \NeX\ \Lyalpha\ lines. This is because the continuum opacity in the cold
stellar winds increases with wavelength. The opacity at the wavelength of the \OVIII\ \Lyalpha\ line
is 27.5 times that at the wavelength of the \SXVI\ \Lyalpha\ line. This outweighs the fact that the
longer wavelength lines originate from the cooler shocked gas far from the line of centres, and so 
travel through less dense regions of the stellar winds.

In general, the inclusion of absorption results in blueward-skewed lines. This is because the gas that
is moving away from the observer is generally on the far side of the system, and so suffers more
absorption. Indeed, for $\theta \ne 0 \degr, 180 \degr$, the velocity shift of the absorbed line 
is always more towards the blue than the corresponding unabsorbed line, though when the system is
viewed from the side of the primary, the lines retain a mean redshift. This is because when the shocked
region is bent away from the observer, there is very little blueshifted emission at all, and so the
fact that redshifted emission is more strongly attenuated than blueshifted emission is less important.
The range of viewing angles over which one expects to see redshifted emission is larger for the
shorter wavelength lines, because absorption is less important for these lines. One would expect
to see redshifted \OVIII\ \Lyalpha\ emission for $\theta \ga 140 \degr$, whereas the \SXVI\ 
\Lyalpha\ line is redshifted when $\theta \ga 90 \degr$.

The HWHMs of the absorbed lines are generally either very similar to those of the corresponding
unabsorbed lines, or considerably smaller (there is very little middle ground). The latter case generally
occurs for the \OVIII\ line when $\theta \approx 45 \degr$--$75 \degr$ or $\theta \ga 90 \degr$,
and for the \NeX\ line when $\theta \ga 90 \degr$ (though the HWHMs of the absorbed and unabsorbed 
lines are again very similar as $\theta$ approaches 180\degr). It also occurs to a lesser extent for the 
\MgXII\ line when $\theta \approx 110 \degr$--$150 \degr$. This narrowing of the absorbed lines occurs
because of the higher opacity at the wavelengths of these lines. The profiles have their redward emission 
strongly attenuated, particularly when $\R > 1$ and the system is being viewed through the dense wind of the 
primary ($\theta \ga 90 \degr$). If it is attenuated to less than half the peak value of the blueward emission, 
only the blueshifted peak will contribute to the HWHM. This results in a considerably smaller HWHM,
though the lines do have large redward tails. However, it is worth noting that in practice such
redward tails may not by observable (they may be lost in the underlying continuum), and so the
dominant observational signature would be a blueshift rather than a broadening.
On the other hand, if the redward emission is not that strongly attenuated, the HWHM will be approximately 
equal to that of the unabsorbed line. This is generally the case for the \SiXIV\ and \SXVI\ lines.

The luminosities of the absorbed lines decrease with $\theta$ for $\R > 1$. This is simply because
as $\theta$ increases, one changes from observing the system through the wind of the secondary to
observing it through the denser wind of the primary.

%%%%%%%%%%%%%%%%%%%%%%%%%%%%%%%%%%%%%%%%%%%%%%%%%%%%%%%%%%%%%%%%%%%%%%%%%%%%%%%%%%%%%%%%%%%%%%%%%%%%%%%%%%%%%%%%%%%%%%%%

\subsection{Lines from systems with different wind parameters but same \R}

\begin{figure*}
  \includegraphics[width=8cm]{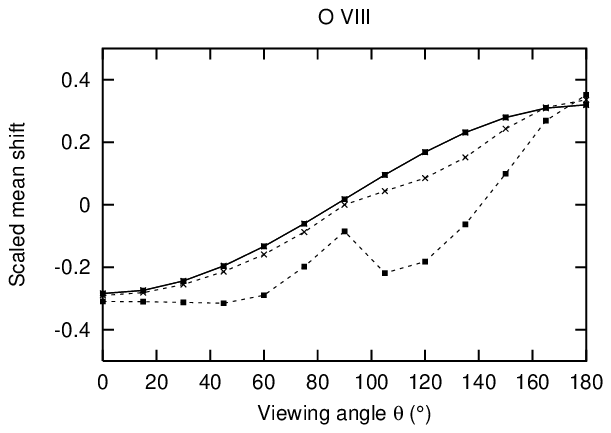}
  \includegraphics[width=8cm]{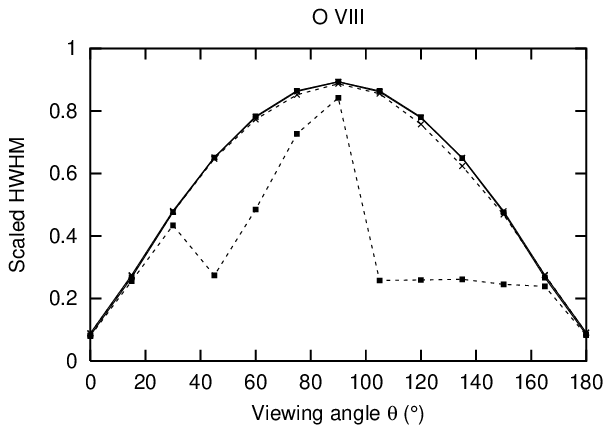}
  \includegraphics[width=8cm]{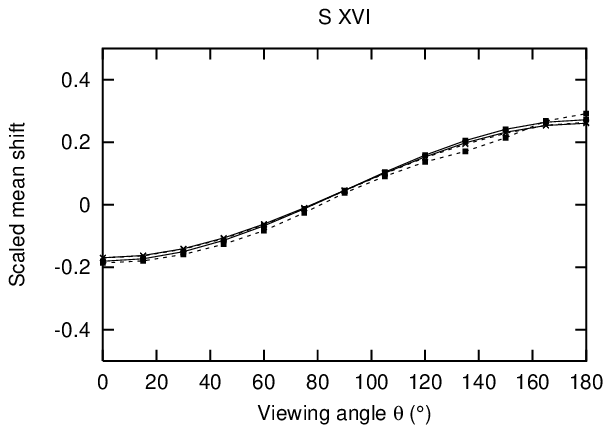}
  \includegraphics[width=8cm]{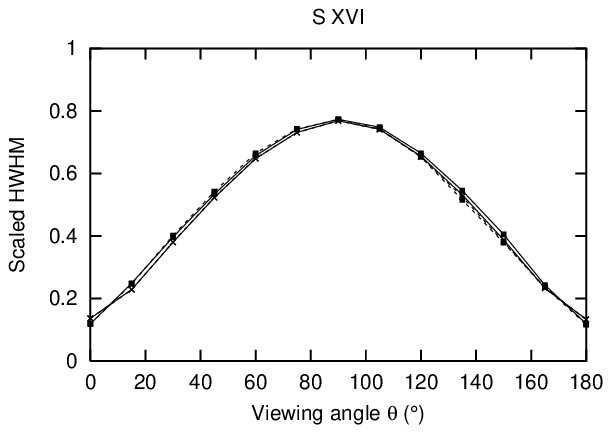}
  \caption{Graphs showing how the mean wavelength shift and HWHM of the unabsorbed (solid)
  and absorbed (dashed) \OVIII\ and \SXVI\ \Lyalpha\ 
  lines vary as a function of viewing angle for two different systems with $\R = \sqrt{5}$.
  The solid squares show results for a system with $\MdotA = 5 \times 10^{-6} \Msolpy$,
  $\MdotB = 1 \times 10^{-6} \Msolpy$ and $\vA = \vB = 2000 \kmps$. The crosses show results for a system 
  with $\MdotA = 5 \times 10^{-7} \Msolpy$, $\MdotB = 1 \times 10^{-7} \Msolpy$ and $\vA = \vB = 2000 \kmps$.}
  \label{fig:cwacwjlinestats}
\end{figure*}

\begin{figure*}
  \includegraphics[width=8cm]{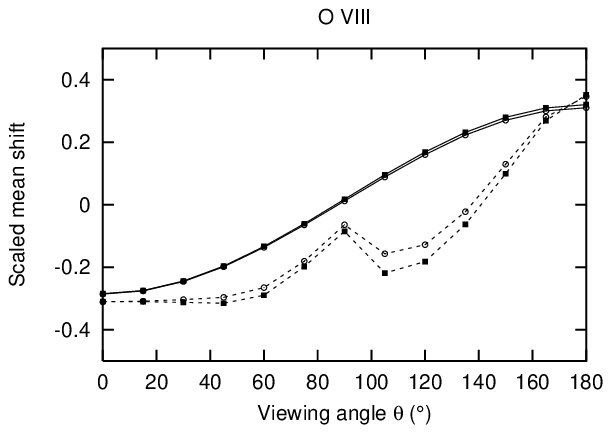}
  \includegraphics[width=8cm]{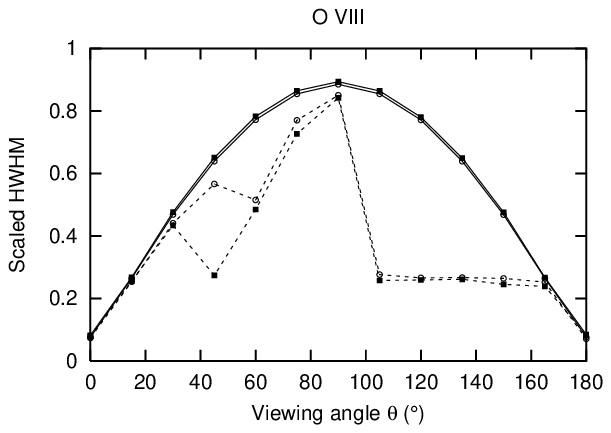}
  \includegraphics[width=8cm]{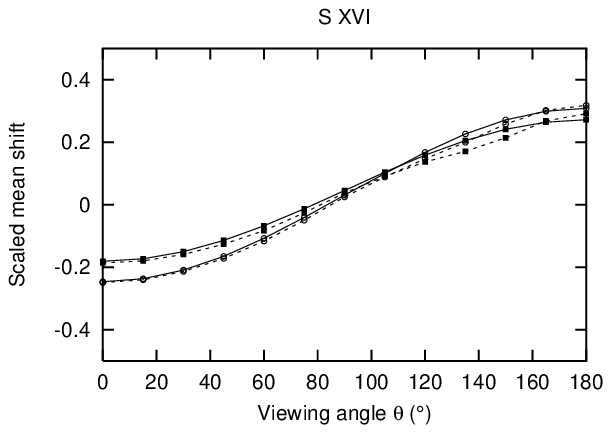}
  \includegraphics[width=8cm]{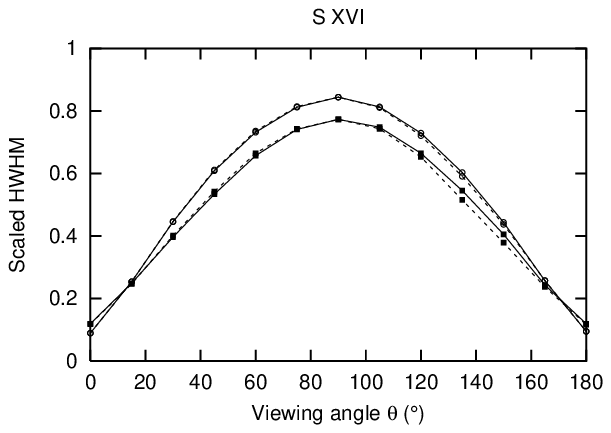}
  \caption{Graphs showing how the mean wavelength shift and HWHM of the unabsorbed (solid) and 
  absorbed (dashed) \OVIII\ and \SXVI\ \Lyalpha\ 
  lines vary as a function of viewing angle for two different systems with $\R = \sqrt{5}$.
  The solid squares show results for a system with $\MdotA = 5 \times 10^{-6} \Msolpy$,
  $\MdotB = 1 \times 10^{-6} \Msolpy$ and $\vA = \vB = 2000 \kmps$. The open circles show results for a system 
  with $\MdotA = 5 \times 10^{-6} \Msolpy$, $\MdotB = 1 \times 10^{-6} \Msolpy$ and $\vA = \vB = 3000 \kmps$.}
  \label{fig:cwjcwmlinestats}
\end{figure*}

Figs.~\ref{fig:cwacwjlinestats} and \ref{fig:cwjcwmlinestats} compare how the mean scaled wavelength
shift and HWHM vary with angle for systems with $\R = \sqrt{5}$ but different wind parameters. Results are shown
for the \OVIII\ and \SXVI\ \Lyalpha\ lines, which are at the two extremes of wavelength studied here.

Fig.~\ref{fig:cwacwjlinestats} compares lines from systems with the same wind speeds but mass-loss rates
that differ by an order of magnitude. For both lines there is no difference in the shapes of the unabsorbed 
profiles between the two systems, though the lines from the system with the larger mass-loss rates
are brighter. However, the system with the higher mass-loss rates gives an absorbed \OVIII\ line that is
narrower and more blueshifted. This is because this system has denser winds, and so the line suffers more attenuation
(especially the redward emission). On the other hand the \SXVI\ line is not as strongly affected by
absorption, and so there is very little difference between the two systems.

For a given \R, the effect of increasing the mass-loss rates on the observed (i.e. absorbed) line luminosities 
is a combination of two competing effects. The intrinsic line luminosity increases as $\Mdot^2$, whereas the 
column density increases as \Mdot. This means that the observed line luminosity scales as
$\Mdot^2 \mathrm{e}^{-\alpha \Mdot}$, where $\alpha$ is a constant that depends on $\theta$ and the opacity 
for the line in question. For the systems described here, we find that the observed line luminosity always
increases with increasing \Mdot\ (for all lines and for all $\theta$). However, for larger mass-loss rates
(few times $10^{-5} \Msolpy$) the increasing absorption starts to dominate, and when $\theta \approx 180 \degr$
(which is when the column density is largest) the observed luminosities of the longer wavelength lines
start to decrease with increasing \Mdot.

Fig.~\ref{fig:cwjcwmlinestats} compares the same two lines from systems with the same mass-loss rates but 
different wind velocities. Once the wavelength shift and the HWHM have been scaled to the wind velocity, there
is very little difference between the two systems. The \OVIII\ line is generally slightly more blueshifted for
the system with the lower wind velocity (since, for the same mass-loss rate, this results in a denser wind and
so more absorption). On the other hand, the wavelength shift of the \SXVI\  line (whether to the blue to the red)
is smaller for the system with the lower wind velocity. This may be because the lower wind speed results in
a lower post-shock temperature, and so the line forms nearer the line of centres for this system. The shocked
region is less bent towards (away from) the observer near the line of centres, resulting in a smaller blueshift
(redshift) as $\theta$ approaches 0\degr\ (180\degr).

%%%%%%%%%%%%%%%%%%%%%%%%%%%%%%%%%%%%%%%%%%%%%%%%%%%%%%%%%%%%%%%%%%%%%%%%%%%%%%%%%%%%%%%%%%%%%%%%%%%%%%%%%%%%%%%%%%%%%%%%

\subsection{Time variability}

Eddies/swirls in the shocked gas are frequently seen in the hydrodynamical simulations of the CWBs used
for this work. These travel from the line of centres to the edge of the grid
(typically $3.4 \times 10^{13} \cm$ from the line of centres) on a time-scale of $\sim 200 \ks$. These swirls may be
due to either instabilities along the contact discontinuity between the two winds \citep{stevens92} or
numerical effects near the line of centres.

In order to investigate whether these swirls have any significant effect on the results, we have compared
line profiles generated from snapshots of the hydrodynamical simulations separated by $\sim 25 \ks$. Over the
course of 10 such snapshots, we find that the calculated wavelength shifts, HWHMs and line luminosities
typically vary by only a few per cent. This suggests
that any observed time variability in the line profiles will be due to changes in the viewing orientation
(because of the orbital motion), rather than to instabilities, etc. in the shocked gas.

%%%%%%%%%%%%%%%%%%%%%%%%%%%%%%%%%%%%%%%%%%%%%%%%%%%%%%%%%%%%%%%%%%%%%%%%%%%%%%%%%%%%%%%%%%%%%%%%%%%%%%%%%%%%%%%%%%%%%%%%

%%%%%%%%%%%%%%%%%%%%%%%%%%%%%%%%%%%%%%%%%%%%%%%%%%%%%%%%%%%%%%%%%%%%%%%%%%%%%%%%%%%%%%%%%%%%%%%%%%%%%%%%%%%%%%%%%%%%%%%%

\subsection{Simulations of observed profiles}

We have carried out some Monte Carlo simulations of observed line profiles from CWBs. Firstly, a
theoretical profile is convolved with a Gaussian, which is a good first order approximation to the
\chandra\ HETGS line response function (\chandra\ Proposers' Observatory 
Guide\footnote{http://asc.harvard.edu/proposer/POG/html/}, ver. 5.0, \S8.2.2). The width of the Gaussian used is 
appropriate for the wavelength of the line in question\footnote{The line resolving power data are taken from 
the file {\tt hetghegD1996-11-01res\_conN0002.rdb}, available from http://space.mit.edu/HETG/caldb/ard\_stat\_hetg.html}.
This convolved profile is then used as a probability distribution function in the Monte Carlo simulations.
For the purposes of these simple calculations, the background/continuum emission in the vicinity of the line
is assumed to negligible.

\begin{figure}
  \includegraphics[width=8cm]{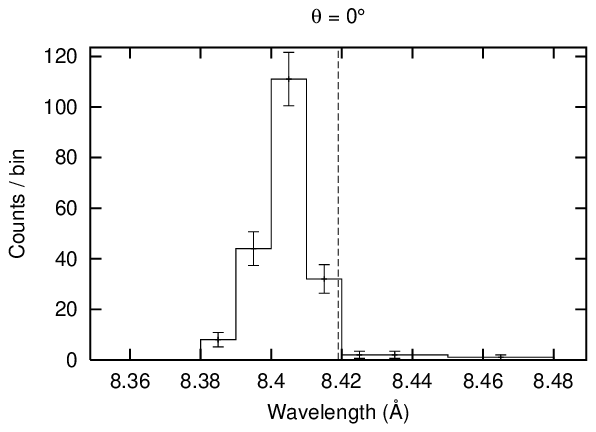}
  \includegraphics[width=8cm]{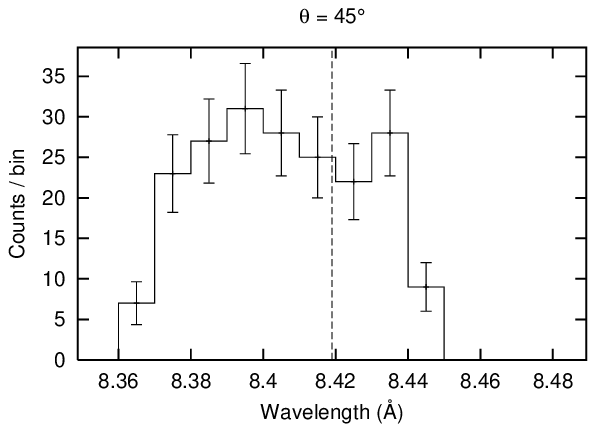}
  \includegraphics[width=8cm]{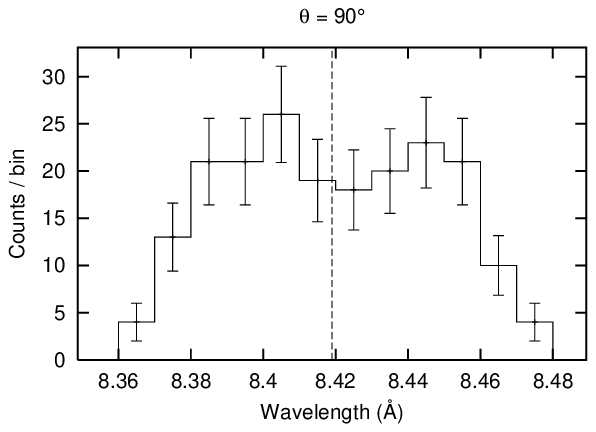}
  \caption{Simulated \MgXII\ \Lyalpha\ profiles as would be observed by the \chandra\ HETGS-HEG. The profiles are for 
  a system with $\R = \sqrt{5}$ ($\MdotA = 5 \times 10^{-6} \Msolpy$, $\MdotB = 1 \times 10^{-6} \Msolpy$,
  $\vA = \vB = 2000 \kmps$) observed from three different viewing angles ($\theta = 0\degr$, 45\degr, 90\degr). 
  There are 200 photons in the line in each case. The dashed vertical line shows the rest wavelength of the brighter
  component of the line (see Table~\ref{table:LineWavelengths}).}
  \label{fig:PhotonSimulation}
\end{figure}

Fig.~\ref{fig:PhotonSimulation} shows the results of simulated observed \MgXII\ \Lyalpha\ line profiles from a 
CWB with $\R = \sqrt{5}$ for three different viewing angles ($\theta = 0\degr$, 45\degr, 90\degr). In each case 
there are assumed to be 200 photons in the line. The results have been binned up to 0.01\angstrom.
These results indicate that the features described in \S\ref{subsec:VariationWithThetaAndR} should easily
be detectable by the \chandra\ HETGS: near conjunction ($\theta = 0 \degr$ or 180\degr) the lines are narrow
with measurable blue- or redshifts, whereas near quadrature ($\theta = 90 \degr$) the profiles are very broad 
and fairly flat-topped with negligible wavelength shifts. Note that for this system $\vA = \vB = 2000 \kmps$,
corresponding to $\Delta \lambda = 0.056 \angstrom$. Hence, near quadrature the HWHM of the observed line is
an appreciable fraction of the wind velocity, as expected. Note also that the photon statistics and the instrumental
response can result in an observed profile that is distinctly different from the theoretical profile for low
counts: compare the middle panel of Fig.~\ref{fig:PhotonSimulation} with the appropriate profile in
Fig.~\ref{fig:mg12lineprofiles}.

%%%%%%%%%%%%%%%%%%%%%%%%%%%%%%%%%%%%%%%%%%%%%%%%%%%%%%%%%%%%%%%%%%%%%%%%%%%%%%%%%%%%%%%%%%%%%%%%%%%%%%%%%%%%%%%%%%%%%%%%

%%%%%%%%%%%%%%%%%%%%%%%%%%%%%%%%%%%%%%%%%%%%%%%%%%%%%%%%%%%%%%%%%%%%%%%%%%%%%%%%%%%%%%%%%%%%%%%%%%%%%%%%%%%%%%%%%%%%%%%%

\section{Discussion}
\label{sec:Discussion}

We have seen that X-ray lines from CWBs show significant variation of wavelength shift and width as a function of
viewing orientation, which is itself a function of orbital phase and inclination. The observation of phase-locked
variation in the X-ray line profiles would therefore provide convincing evidence for X-ray emission from a 
colliding wind region. In this respect, an eclipsing binary would be the most interesting system,
since during the course of its orbit the full range of viewing angles discussed above would be observed. In contrast,
a system with an inclination of 0\degr\ would always be observed at $\theta = 90\degr$. Any phase-locked variation
in this case would just be associated with a change in orbital separation and hence the density of the shocked gas.
This would probably just manifest itself as a change in line luminosity, not as a change in line shift or width.

Line profiles that vary during the orbit of a CWB have already been observed in the \chandra\ spectra of WR 140
(Pollock et al., in prep.). The first observation was taken just before periastron at an orbital phase of 1.982 
according to the ephemeris of \citet{williams90}. At this phase the wind collision region is bent approximately 
towards the observer (see \citet{white95} for a diagram of WR 140's orbit), and the lines are significantly 
blueshifted, as would be expected. The second observation was taken just after periastron at a phase of 2.027. 
Although there are considerably fewer counts in this second spectrum, the lines are clearly much broader, and there
is also some evidence of redshifts. While we have not compared in detail our theoretical profiles with the spectra
of WR 140, these observations seem to provide at least a qualitative confirmation of some of our general predictions.

A more detailed comparison of our model line profiles with observed high-resolution X-ray spectra of CWBs may enable
us to place constraints on certain parameters of the system. In practice, one would just fit to the shifts
and widths of the lines, and scale the luminosities of the theoretical lines to the observed luminosities. (This approach
of fitting the line \textit{shapes}, not the luminosities, has already been used by \citet*{kramer03a} to model 
the line emission from the single O star $\zeta$ Pup.) For a CWB, the main parameters would be the wind momentum ratio 
\R\ and the viewing angle $\theta$, although it may also be possible to place constraints on the individual wind parameters.
For this method to be effective, one would want to fit as many lines as possible simultaneously. While we have only
presented results for \Lyalpha\ lines here, the method described could be used for any line that is not affected
by the density (e.g. \Lybeta\ and \Lygamma\ lines, and the resonant component of the \textit{fir} triplets from 
He-like ions).

The model presented here for calculating X-ray line profiles from CWBs is fairly simple, but it contains the 
essential physics and gives an overview of how the profiles vary with wind parameters and viewing angle.
When the model comes to be applied to real systems, extra physics will be included in the calculations as is 
appropriate for the systems in question. This could include modifying
the hydrodynamical simulations (e.g. radiative cooling for systems with small orbital separations and/or large mass-loss 
rates \citep{stevens92} or radiative driving for close systems \citep{pittard98b}), or modifying the actual line
profile calculations (e.g. taking into account non-solar abundances for WR+O binaries).

%%%%%%%%%%%%%%%%%%%%%%%%%%%%%%%%%%%%%%%%%%%%%%%%%%%%%%%%%%%%%%%%%%%%%%%%%%%%%%%%%%%%%%%%%%%%%%%%%%%%%%%%%%%%%%%%%%%%%%%%

\section{Summary}
\label{sec:Summary}

We have presented theoretical X-ray line profiles from model colliding wind systems for a range of mass-loss rates,
wind velocities and viewing angles. A wide range of profile shapes is possible, varying with orbital inclination and 
phase.

For a high inclination binary with $\R > 1$ there is a general blue/redshift at phases 0 and 0.5 (i.e. at the two
conjunctions). The profiles are narrow ($\mathrm{HWHM} \sim 0.1 v_\infty$) and approximately Gaussian. By contrast, at
quadrature the profiles are generally very broad ($\mathrm{HWHM} \sim v_\infty$), flat-topped and unshifted. Local 
absorption has a major effect on the observed profiles. When a system with relatively large mass-loss rates
(few times $10^{-6} \Msolpy$) is viewed through the dense wind of the primary, the lines are generally skewed 
towards the blue. This is because the redshifted emission is strongly attenuated. This effect is most noticeable 
for the lower energy lines ($E \la 1 \kev$). These blueward-skewed profiles generally have large redward tails. In
practice these tails may not be observable, and so the line would be seen as fairly narrow and blueshifted, rather
than broad and blueward-skewed.

A binary with a low orbital inclination will exhibit less orbital variation of line width and shift, the extreme case
being a binary with $i = 0\degr$, for which we would expect no orbital variation in the line profile.

After the line shifts and widths have been scaled to the wind velocity, there is very little difference between the 
results from systems with $\vA = \vB = 2000 \kmps$ and with $\vA = \vB = 3000 \kmps$. The former systems have denser 
winds for the same mass-loss rate, and hence absorption has a larger effect. However, it does not significantly affect 
the results.

Simple Monte Carlo simulations of observed X-ray line profiles show that the features discussed above will be
detectable in good \chandra\ HETGS spectra of CWBs. Comparing our theoretical profiles with such spectra 
potentially offers another method for determining the wind parameters of CWBs, and may also give us new insights
into the structure of the wind-wind collision regions.

%%%%%%%%%%%%%%%%%%%%%%%%%%%%%%%%%%%%%%%%%%%%%%%%%%%%%%%%%%%%%%%%%%%%%%%%%%%%%%%%%%%%%%%%%%%%%%%%%%%%%%%%%%%%%%%%%%%%%%%%

\section*{Acknowledgments}

We gratefully acknowledge funding from the School of Physics \& Astronomy (DBH) and PPARC (IRS, JMP).

%%%%%%%%%%%%%%%%%%%%%%%%%%%%%%%%%%%%%%%%%%%%%%%%%%%%%%%%%%%%%%%%%%%%%%%%%%%%%%%%%%%%%%%%%%%%%%%%%%%%%%%%%%%%%%%%%%%%%%%%

%%%%%%%%%%%%%%%%%%%%%%%%%%%%%%%%%%%%%%%%%%%%%%%%%%%%%%%%%%%%%%%%%%%%%%%%%%%%%%%%%%%%%%%%%%%%%%%%%%%%%%%%%%%%%%%%%%%%%%%%


\begin{thebibliography}{}

\bibitem[\protect\citeauthoryear{Cassinelli, Miller, Waldron, MacFarlane \&
  Cohen}{Cassinelli et~al.}{2001}]{cassinelli01a}
Cassinelli J.~P.,  Miller N.~A.,  Waldron W.~L.,  MacFarlane J.~J.,    Cohen
  D.~H.,  2001, ApJ, 554, L55

\bibitem[\protect\citeauthoryear{Cherepashchuk}{Cherepashchuk}{1976}]{cherepas%
hchuk76}
Cherepashchuk A.~M.,  1976, SvA Lett., 2, 138

\bibitem[\protect\citeauthoryear{Chlebowski}{Chlebowski}{1989}]{chlebowski89b}
Chlebowski T.,  1989, ApJ, 342, 1091

\bibitem[\protect\citeauthoryear{Cohen, de Messi{\`e}res, MacFarlane, Miller,
  Cassinelli, Owocki \& Liedahl}{Cohen et~al.}{2003}]{cohen03}
Cohen D.~H.,  de Messi{\`e}res G.~E.,  MacFarlane J.~J.,  Miller N.~A.,
  Cassinelli J.~P.,  Owocki S.~P.,    Liedahl D.~A.,  2003, ApJ, 586, 495

\bibitem[\protect\citeauthoryear{Colella \& Woodward}{Colella \&
  Woodward}{1984}]{colella84}
Colella P.,  Woodward P.~R.,  1984, J. Comput. Phys., 54, 174

\bibitem[\protect\citeauthoryear{Corcoran, Swank, Petre, Ishibashi, Davidson,
  Townsley, Smith, White, Viotti \& Damineli}{Corcoran
  et~al.}{2001}]{corcoran01b}
Corcoran M.~F. et al., 2001, ApJ, 562, 1031

\bibitem[\protect\citeauthoryear{Feldmeier, Puls \& Pauldrach}{Feldmeier
  et~al.}{1997}]{feldmeier97a}
Feldmeier A.,  Puls J.,    Pauldrach A.~W.~A.,  1997, A\&A, 322, 878

\bibitem[\protect\citeauthoryear{Ignace}{Ignace}{2001}]{ignace01a}
Ignace R.,  2001, ApJ, 549, L199

\bibitem[\protect\citeauthoryear{Kahn, Leutenegger, Cottam, Rauw, Vreux, den
  Boggende, Mewe \& G{\"u}del}{Kahn et~al.}{2001}]{kahn01}
Kahn S.~M.,  Leutenegger M.~A.,  Cottam J.,  Rauw G.,  Vreux J.-M.,  den
  Boggende A.~J.~F.,  Mewe R.,    G{\"u}del M.,  2001, A\&A, 365, L312

\bibitem[\protect\citeauthoryear{Kramer, Cohen \& Owocki}{Kramer
  et~al.}{2003}]{kramer03a}
Kramer R.~H.,  Cohen D.~H.,    Owocki S.~P.,  2003, ApJL, submitted
  (astro-ph/0211550)

\bibitem[\protect\citeauthoryear{Leutenegger, Kahn \& Ramsay}{Leutenegger
  et~al.}{2003}]{leutenegger03}
Leutenegger M.~A.,  Kahn S.~M.,    Ramsay G.,  2003, ApJ, 585, 1015

\bibitem[\protect\citeauthoryear{Lucy}{Lucy}{1982}]{lucy82}
Lucy L.~B.,  1982, ApJ, 255, 286

\bibitem[\protect\citeauthoryear{Lucy \& White}{Lucy \& White}{1980}]{lucy80}
Lucy L.~B.,  White R.~L.,  1980, ApJ, 241, 300

\bibitem[\protect\citeauthoryear{Luo, McCray \& Mac~Low}{Luo
  et~al.}{1990}]{luo90}
Luo D.,  McCray R.,    Mac~Low M.-M.,  1990, ApJ, 362, 267

\bibitem[\protect\citeauthoryear{Miller, Cassinelli, Waldron, MacFarlane \&
  Cohen}{Miller et~al.}{2002}]{miller02b}
Miller N.~A.,  Cassinelli J.~P.,  Waldron W.~L.,  MacFarlane J.~J.,    Cohen
  D.~H.,  2002, ApJ, 577, 951

\bibitem[\protect\citeauthoryear{Owocki, Castor \& Rybicki}{Owocki
  et~al.}{1988}]{owocki88}
Owocki S.~P.,  Castor J.~I.,    Rybicki G.~B.,  1988, ApJ, 335, 914

\bibitem[\protect\citeauthoryear{Owocki \& Cohen}{Owocki \&
  Cohen}{2001}]{owocki01}
Owocki S.~P.,  Cohen D.~H.,  2001, ApJ, 559, 1108

\bibitem[\protect\citeauthoryear{Pittard}{Pittard}{1998}]{pittard98b}
Pittard J.~M.,  1998, MNRAS, 300, 479

\bibitem[\protect\citeauthoryear{Pittard \& Corcoran}{Pittard \&
  Corcoran}{2002}]{pittard02a}
Pittard J.~M.,  Corcoran M.~F.,  2002, A\&A, 383, 636

\bibitem[\protect\citeauthoryear{Pittard \& Stevens}{Pittard \&
  Stevens}{1997}]{pittard97}
Pittard J.~M.,  Stevens I.~R.,  1997, MNRAS, 292, 298

\bibitem[\protect\citeauthoryear{Pollock}{Pollock}{1987}]{pollock87}
Pollock A.~M.~T.,  1987, ApJ, 320, 283

\bibitem[\protect\citeauthoryear{Prilutskii \& Usov}{Prilutskii \&
  Usov}{1976}]{prilutskii76}
Prilutskii O.~F.,  Usov V.~V.,  1976, SvA, 20, 2

\bibitem[\protect\citeauthoryear{Schulz, Canizares, Huenemoerder \& Lee}{Schulz
  et~al.}{2000}]{schulz00a}
Schulz N.~S.,  Canizares C.~R.,  Huenemoerder D.,    Lee J.~C.,  2000, ApJ,
  545, L135

\bibitem[\protect\citeauthoryear{Skinner, G{\"u}del, Schmutz \&
  Stevens}{Skinner et~al.}{2001}]{skinner01}
Skinner S.~L.,  G{\"u}del M.,  Schmutz W.,    Stevens I.~R.,  2001, ApJ, 558,
  L113

\bibitem[\protect\citeauthoryear{Stevens, Blondin \& Pollock}{Stevens
  et~al.}{1992}]{stevens92}
Stevens I.~R.,  Blondin J.~M.,    Pollock A.~M.~T.,  1992, ApJ, 386, 265

\bibitem[\protect\citeauthoryear{Stevens, Corcoran, Willis, Skinner, Pollock,
  Nagase \& Koyama}{Stevens et~al.}{1996}]{stevens96}
Stevens I.~R.,  Corcoran M.~F.,  Willis A.~J.,  Skinner S.~L.,  Pollock
  A.~M.~T.,  Nagase F.,    Koyama K.,  1996, MNRAS, 283, 589

\bibitem[\protect\citeauthoryear{Waldron \& Cassinelli}{Waldron \&
  Cassinelli}{2001}]{waldron01}
Waldron W.~L.,  Cassinelli J.~P.,  2001, ApJ, 548, L45

\bibitem[\protect\citeauthoryear{White \& Becker}{White \&
  Becker}{1995}]{white95}
White R.~L.,  Becker R.~H.,  1995, ApJ, 451, 352

\bibitem[\protect\citeauthoryear{Williams, van~der Hucht, Pollock, Florkowski,
  van~der Woerd \& Wamsteker}{Williams et~al.}{1990}]{williams90}
Williams P.~M.,  van~der Hucht K.~A.,  Pollock A.~M.~T.,  Florkowski D.~R.,
  van~der Woerd H.,    Wamsteker W.~M.,  1990, MNRAS, 243, 662

\end{thebibliography}
\end{document}